%% file: main.tex
\documentclass[journal]{vgtc}                     % final (journal style)
\onlineid{1499}

\vgtccategory{Research}

\title{Not Always Top-Left: Untangling the Signals that Guide\\ Dashboard Reading Order}

\author{%
  \authororcid{Nicole Sultanum}{0000-0001-8608-1427},
  \authororcid{Vidya Setlur}{0000-0003-3722-406X}
}

\authorfooter{
  \item
  	Nicole Sultanum and Vidya Setlur are with Tableau Research.
  	E-mail: \{nsultanum, vsetlur\}@tableau.com
}

\input{tex/00_abstract}

\teaser{
  \centering
  \includegraphics[width=\linewidth]{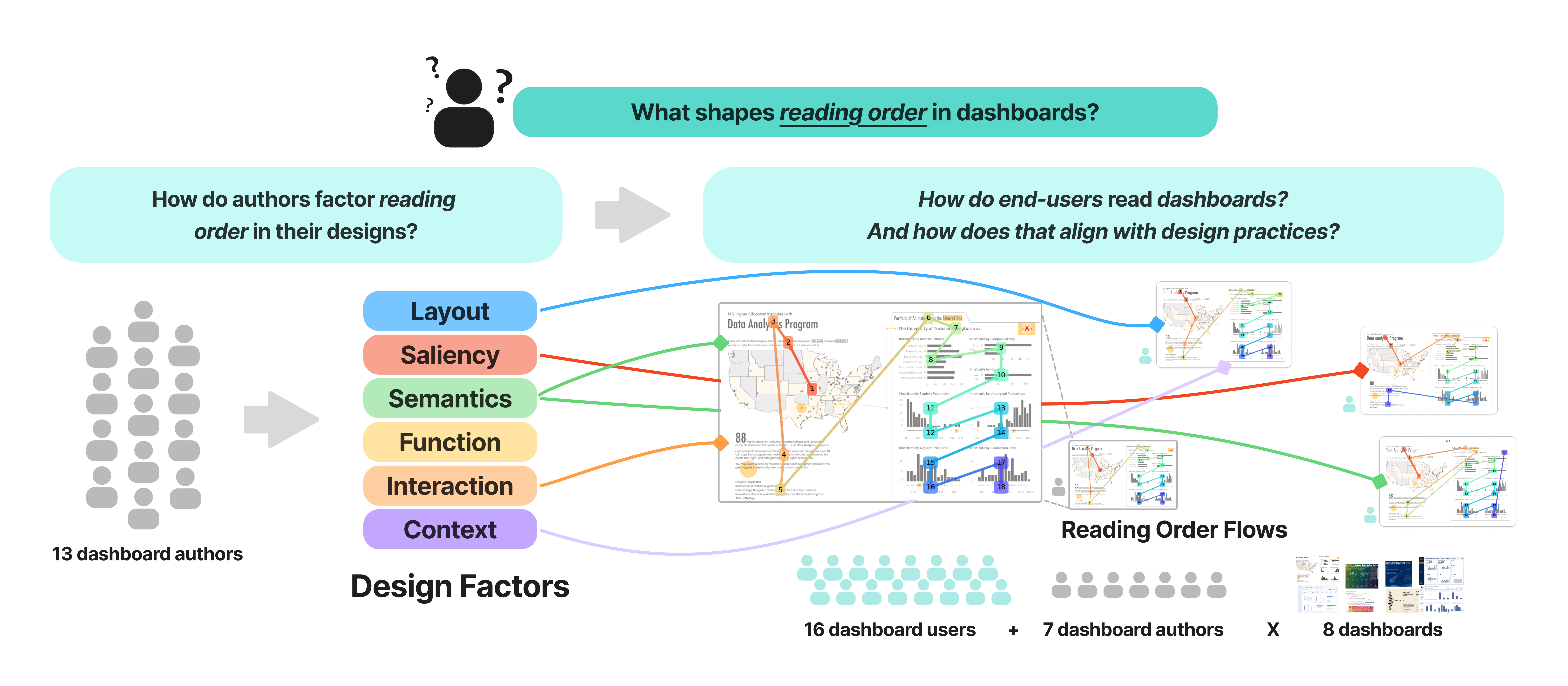}
  \caption{%
    We investigate how users \textit{read} dashboards, i.e., how they \textit{traverse them for understanding}, and how authors design with these expectations in mind. Through interviews with 13 dashboard authors, we identify six key \textit{design factors} that shape reading order: \textit{Layout}, \textit{Visual Saliency}, \textit{Semantics}, \textit{Function}, \textit{Interaction}, and \textit{User Context}. We then conduct a study with 16 users across 8 dashboards, alongside 7 corresponding authors, to elicit reading flow instances and rationales. Comparing these perspectives reveals where user behavior aligns with or diverges from author intent, providing a clearer understanding of dashboard reading order.
  }
  \label{fig:teaser}
}

\graphicspath{{figs/}{figures/}{pictures/}{images/}{./}} % where to search for the images

\usepackage{tabu}                      % only used for the table example
\usepackage{booktabs}                  % only used for the table example
\usepackage{lipsum}                    % used to generate placeholder text
\usepackage{mwe}                       % used to generate placeholder figures
\usepackage{ccicons}                   % package to be able to use icons from creative commons

\usepackage{mathptmx}                  % use matching math font

\include{tex/custom-imports-and-commands}
\begin{document}

%%%%%%%%%%%%%%%%%%%%%%%%%%%%%%%%%%%%%%%%%%%%%%%%%%%%%%%%%%%%%%%%
%%%%%%%%%%%%%%%%%%%%%% START OF THE PAPER %%%%%%%%%%%%%%%%%%%%%%
%%%%%%%%%%%%%%%%%%%%%%%%%%%%%%%%%%%%%%%%%%%%%%%%%%%%%%%%%%%%%%%%

%% The ``\maketitle'' command must be the first command after the
%% ``\begin{document}'' command. It prepares and prints the title block.
%% the only exception to this rule is the \firstsection command
\firstsection{Introduction}

\maketitle

\input{tex/01-intro}
\input{tex/02-rw}

\input{tex/03-formative}
\input{tex/04-author-user-studies}
\input{tex/05-findings}

\input{tex/06-reflections}

\input{tex/07-limitations}
\input{tex/08-conclusion}

%% if specified like this the section will be omitted in review mode
% \acknowledgments{%
% }

\newpage

\bibliographystyle{abbrv-doi-hyperref}

\bibliography{references}

\appendix % You can use the `hideappendix` class option to skip everything after \appendix
\crefalias{section}{appendix} % this is to make sure that cleverref switches to referring to Appx. X from here on

\newpage

\section{Appendix: Additional Reading Flow Patterns}
\input{tex/09-appendix}

\end{document}

%% file: tex/00_abstract.tex
\abstract{Dashboards are widely used interfaces for data analysis, combining multiple visualizations, text, and interactive controls within a single view. While dashboard authors often structure layouts to suggest a logical consumption flow, users may interpret and navigate dashboards differently depending on the interplay between design features, analytical goals, and personal preferences. In this work, we investigate how people make sense of dashboards by examining their \textit{reading orders}, i.e., the sequences in which users engage with dashboard components. We conduct a mixed-methods study with 18 dashboard authors and 16 end-users, capturing how participants \rr{design for and} reason through these component \rr{transitions}. Through qualitative and quantitative analyses of participant-generated flows, we outline a set of factors that influence dashboard reading order, including layout, visual saliency, semantics, functional roles, interaction, and user context. We also identify emergent reading patterns and analyze them through aggregate and variability measures, revealing where users converge and diverge in their interpretations. Finally, we discuss implications and opportunities for computational approaches that aim to automatically model, guide, or serialize dashboard consumption.
} 

%% file: tex/custom-imports-and-commands.tex
\usepackage{multicol}
\usepackage{soul}

\usepackage{amsmath} 
\usepackage{algorithm}
\usepackage{algpseudocode}
\usepackage{comment}

\usepackage{changepage}
\usepackage{tcolorbox}
\usepackage{subcaption}
\usepackage{url}
\usepackage{xurl} 

\usepackage{tikz}
\usepackage{fontawesome5}

\usepackage{xcolor}
\definecolor{newred}{RGB} {230, 70, 40} 

\newcommand{\vidya}[1]{\textcolor{purple}{VS: #1}}
\newcommand{\nicole}[1]{\textcolor{red}{(NS: #1)}}
\newcommand{\todo}[1]{\textcolor{red}{(#1)}}

\newcommand{\pid}[1]{{$#1$}}
\newcommand{\db}[1]{{$D{#1}$}}

\newcommand{\pheading}[1]{\vspace{4px}\noindent\textbf{#1}}

\newcommand{\qte}[2]{\emph{``{#1}''} (\pid{#2})}

\newcommand{\implication}[2]{
\begin{adjustwidth}{1em}{}
\textbf{Design Implication {#1}:} {\textit{#2}}
\end{adjustwidth}
}

\newenvironment{tight_itemize}{\begin{itemize} \itemsep
-1.5pt}{\end{itemize}}

\newcommand{\rr}[1]{\textcolor{black}{#1}}
\newcommand{\cre}[1]{\textcolor{black}{#1}}

\newcommand{\coloredunderline}[2][black]{\setulcolor{#1}% Set the underline color (defaults to red)
\setul{0.2ex}{1.5pt}\ul{#2}% Apply the underline
}

\newcommand{\theme}[2]{\coloredunderline[#1]{#2}}

\definecolor{thclayout}{RGB}{105, 202, 255}%#69caff
\definecolor{thcsaliency}{RGB}{237, 154, 195}%#ed9ac3
\definecolor{thcsemantic}{RGB}{132, 213, 152}%#84d598
\definecolor{thcinteract}{RGB}{247, 184, 129}%#f7b881
\definecolor{thcfunction}{RGB}{227, 219, 56}%#e3db38
\definecolor{thccontext}{RGB}{193, 164, 241}%#c1a4f1

\newcommand{\thlayout}[1]{\theme{thclayout}{#1}}

\newcommand{\thsaliency}[1]{\theme{thcsaliency}{#1}}

\newcommand{\thsemantic}[1]{\theme{thcsemantic}{#1}}

\newcommand{\thinteract}[1]{\theme{thcinteract}{#1}}

\newcommand{\thfunction}[1]{\theme{thcfunction}{#1}}

\newcommand{\thcontext}[1]{\theme{thccontext}{#1}}

%% file: tex/01-intro.tex
\label{sec:intro}
Multi-view dashboards combine text, charts, and interactive components into a unified interface for exploring and communicating data~\cite{bach2022dashboard}. Their visual structure expressed through layout, hierarchy, and composition reflects deliberate design choices intended to guide interpretation and support efficient understanding of the data being presented~\cite{plume, srinivasan2023azimuth}. In practice, these designs often imply a \textit{logical flow of information}, suggesting how users should navigate and interpret the dashboard.

However, how users actually traverse dashboards is far less understood. Prior work has examined dashboard design through typologies, layout structure, and component relationships~\cite{bach2022dashboard, srinivasan2024dashboard,zeng2023semi}, as well as through studies of visual attention and scanning behavior~\cite{nielsen2006fpattern, itti:1998, bylinski:eyefixation:2017, yang2025dashboard}. Yet these lines of work largely consider structure or attention in isolation, offering limited insight into 
%how users determine the \textit{sequence} 
\rr{how design and contextual factors influence the \textit{sequence}}
in which they engage with dashboard components during data exploration\cre{, and \textit{why}}. We argue that understanding these sequences is critical for moving beyond static representations of dashboards \rr{and toward models that better capture usage} in practice. %We argue that understanding these sequences is critical for moving beyond static representations of dashboards to better capture how they are used in practice.  

In this paper, we take initial steps toward characterizing reading order for dashboards. We refer to \textit{reading flow} as the ordered (and often non-linear) progression through which users attend to, interpret, and relate dashboard elements for understanding.
To study this process empirically, we conducted two studies. First, we ran an interview-based formative study with 13 dashboard authors to learn how aspects of reading order are taken into account in their dashboard design process. These findings informed a list of six design factors and five design principles reported to influence dashboard reading flows. To validate and expand on these findings, we then conducted a mixed-methods study with 16 end-users and 7 dashboard authors (including 2 from the formative study) to collect reading flow \textit{instances}, i.e., explicit, user-specified sequences of component engagement that operationalize reading flow for a given dashboard. 
Insights from reading flow instances and their rationales revealed nine reading flow patterns with implications \rr{for reading} order design. They also highlight opportunities for future research and applications, leveraging semantically aware representations to better support dashboard authoring and consumption.

We claim the following research contributions: 
\begin{tight_itemize}
\item Findings from two empirical studies of dashboard reading order that reveal how users sequence components based on dashboard features, and characterize where and why user-driven reading flows diverge from author-intended structure.

\item A structured set of dashboard design \textit{factors} that account for spatial, semantic, and relational properties of dashboard elements that affect reading order, and recurring reading flow \textit{patterns} that represent common traversal strategies.

\item Design implications and future opportunities for dashboard design and tooling, including layout-aware authoring, accessibility improvements, onboarding support, and reading-order-aware intelligent systems.
\end{tight_itemize}

%% file: tex/02-rw.tex
\section{Related Work}
\label{sec:related-work}

Understanding dashboard reading order requires drawing from multiple lines of research that examine how visual interfaces such as dashboards are structured and perceived. We therefore review related research spanning (1) dashboard design \& layout, (2) visual attention \& document reading order, and (3) multiview understanding.

\subsection{Dashboard Design and Layout}
A substantial body of work has examined how dashboards are structured and designed. Bach et al.~\cite{bach2022dashboard} characterize dashboard design as inherently multi-dimensional, involving tradeoffs between abstraction, screen space, interaction, and information density. Complementing this, Srinivasan et al.~\cite{srinivasan2024dashboard} analyze large collections of dashboards, modeling them as graphs of content blocks and relationships to uncover common design patterns, while Zeng et al.~\cite{zeng2023semi} explore semi-automated approaches to layout generation.

Dashboards span a broad design space across dimensions such as purpose, audience, interactivity, and data semantics, ranging from glanceable monitoring interfaces to narrative visualizations~\cite{sarikaya:2019}. This diversity underscores that layout and organization are closely tied to intended use. Prior work has also examined how dashboards encode hierarchy and relationships between components. Systems such as Azimuth~\cite{srinivasan2023azimuth} and practitioner guidelines emphasize hierarchical organization, grouping, and structured representations to support navigation. Lin et al.~\cite{lin:2024} further show that layout can be partially explained through learned mappings between view properties and spatial placement, highlighting the role of both individual components and pairwise relationships in shaping structure.

Beyond static layout, dashboards have been framed as supporting an analytical “conversation” between users and systems. Setlur et al.~\cite{setlur2023heuristics} propose conversational design heuristics that guide iterative exploration, while systems such as BOLT~\cite{srinivasan2023bolt}, DashBot~\cite{deng2022dashbot}, and MEDLEY~\cite{pandey2023medley} support dashboard authoring through intent-driven, sequential, or goal-oriented generation. These approaches highlight that dashboards are organized around analytical goals and evolving interactions rather than fixed arrangements of charts.

This body of work focuses on how dashboards are structured or generated, rather than how users traverse and interpret them. Our work examines dashboard reading order and how users sequentially engage with components while making sense of the information presented.

\subsection{Visual Attention and Document Reading Order}
Research in visual attention provides a foundation for understanding how users prioritize elements in interfaces. Eye-tracking studies show that users often follow predictable scanning patterns, such as F- or Z-shaped behaviors~\cite{nielsen2006fpattern}, while saliency models demonstrate that attention is driven by low-level features such as color, contrast, and size~\cite{itti:1998}.

More recent work considers layout as an ordered visual organization that shapes perception and behavior. Zhang et al.~\cite{zhang2024effects} show that more structured layouts improve search efficiency and reduce perceived complexity. Visual \rr{saliency} further influences interpretation: embellishments affect engagement and recall~\cite{bateman:2010}, and memorability is linked to visual features~\cite{borkin:2013}. Eye-tracking analyses using fixation-based metrics \rr{model} how users allocate attention and transition between \rr{visualization elements}~\cite{bylinski:eyefixation:2017}, while \rr{eye-tracking} studies of dashboards reveal systematic patterns such as hierarchical scanning, upper-left spatial bias, \rr{and early fixation on text elements}~\cite{yang2025dashboard}.
Attention is also shaped by user characteristics, including cognitive abilities and task context~\cite{toker:2013}.
\rr{These findings confirm the effect of various dashboard design factors on scanning patterns and provide partial signals on sequencing patterns~\cite{yang2025dashboard}, but do not inform how these attention behaviors map to reasoning patterns when navigating dashboards; i.e., user intent.} 

A complementary line of work examines reading order in documents, where the goal is to infer or model the sequence in which content should be read. Early work in document analysis formalizes reading order as a structural problem, focusing on recovering logical sequences from spatial layouts, particularly in complex documents such as newspapers or scanned pages~\cite{ferilli2014abstract}. Research on reading behavior in digital documents and hypertext shows that users often follow non-linear and goal-driven navigation strategies. Studies of web document reading indicate that readers selectively prioritize regions based on perceived importance, layout cues, and task goals, rather than strictly following linear order~\cite{buscher:2010,SALMERON20111143}. Work on passage-level reading behavior further demonstrates that users dynamically shift attention across sections, often skipping or revisiting content based on relevance~\cite{wu2023passage}.

Text plays a central role in guiding attention, often serving as an early anchor through titles, labels, and legends~\cite{matzen:2017}. Attention evolves over time, with shifts driven by task relevance, complexity, and cognitive state~\cite{arunkumar:2025}. Beyond visual features, structural properties such as visualization type and data characteristics influence perception~\cite{listructureaware:2022}, and interaction mechanisms can alter attention through changes in cognitive load~\cite{schmid2023influence}. Interpretation also operates across semantic abstraction, from low-level encodings to higher-level insights~\cite{Lundgard2021AccessibleVV}. \rr{These studies show that reading order emerges from an interplay of spatial layout, semantic content, and user goals. However, they primarily examine linear documents, which differ fundamentally from dashboards and their complex interplay of visual, textual, and interactive design features to support various analytical goals. As such, findings for document reading provide only partial insight into how users peruse dashboards.}
%Our work extends beyond predicting reading sequences by characterizing how reading flows emerge during dashboard sensemaking and how they are shaped by layout, semantics, interaction, and user intent.} %More recent approaches leverage machine learning to jointly model text and layout for reading order detection, demonstrating that both spatial structure and semantic content are necessary to predict \rr{document} reading sequences~\cite{soto2019visual,zhang2020trie}. \vidya{From nicole: emphasize works in this paragraph are all about documents, and their insights are only partially applicable to dashboards, so there's still a research gap here?}.

These findings suggest that \rr{dashboard} attention is shaped by an interplay of visual, structural, semantic, and user-dependent factors. However, prior work primarily models where users look \rr{and what draws their attention~\cite{yang2025dashboard, bylinski:eyefixation:2017, zhang2024effects}}, \rr{but provides limited insight on how they \textit{transition} attention across elements and how these transitions map to dashboard sensemaking}. While layout and linearity influence reading behavior~\cite{wang2021layoutreader, peitek2020what, yang2025dashboard}, they do not fully capture how users traverse complex, multi-component dashboards. Our work shifts focus from static attention to reading order, examining how users sequentially engage with dashboard elements for understanding.

\subsection{Multiview Understanding}
Dashboards can be seen as multi-view visualizations, where multiple coordinated views support complex analytical tasks. Work on dashboard repositories shows that users often engage in information-seeking by browsing and comparing related dashboards, relying on semantic similarity to identify relevant content~\cite{oppermann:2021}. These behaviors highlight the importance of understanding how users navigate both across and within dashboards while making sense of the information.

Understanding is inherently iterative and non-linear, with users constructing meaning through cycles of exploration and refinement~\cite{pirolli2005sensemaking}. Systems such as DashGuide~\cite{dashguide} \cre{capture user transitions} between views \cre{for analysis and storytelling}, highlighting the role of sequencing and flow. Research on coordinated multiple views similarly shows how users navigate across views to explore relationships and refine hypotheses~\cite{roberts:2007}, \cre{with transitions shaped by} system design and user intent.

Design principles for multi-view visualizations focus on supporting interpretation and integration. Maintaining consistency in encodings, scales, and ordering facilitates comparison and reduces cognitive effort, while inconsistencies can introduce confusion~\cite{qu:2018}. At the same time, relationships between views can be modeled through context-aware approaches such as Chart2Vec, which captures co-occurrence and sequential dependencies between charts, treating dashboards as structured collections rather than independent views~\cite{chart2vec}. Multi-view visualizations also introduce boundaries across data, representation, and semantics that can hinder synthesis, requiring users to bridge gaps; techniques such as linking, highlighting, \rr{morphing}, and embedding help integrate information into a coherent sensemaking space\rr{~\cite{salako2025animated, sun2023boundary}}.

While prior work emphasizes transitions, consistency, and relationships between views, our work shifts focus to reading order, examining how users sequence their engagement with dashboard components to understand the underlying data and insights.

%% file: tex/03-formative.tex
\section{Designing for Dashboard Reading Order}
\label{sec:interviews}

While prior work examined dashboard layout\rr{~\cite{bach2022dashboard,srinivasan2023bolt,lin:2024,zeng2023semi}}, visual attention\rr{~\cite{yang2025dashboard,zhang2024effects,bylinski:eyefixation:2017,itti:1998}}, and multi-view analysis\rr{~\cite{roberts:2007,qu:2018,chart2vec}} in isolation, there \rr{is} limited understanding of how these factors shape the ways in which users sequentially engage with dashboard content. In particular, the notion of ``reading order'' in dashboards remains implicit, often inferred from layout conventions but rarely examined as a distinct aspect of user behavior and understanding. 
To better ground this concept in real-world practice, we conducted formative interviews with expert dashboard creators. Our goal was to develop an initial, practice-informed understanding of dashboard reading order, including the factors practitioners consider when structuring the flow of information and how these considerations manifest in their design process. 

\subsection{Methodology}
We recruited 13 professional dashboard practitioners (\pid{P1}-\pid{P13}) from relevant online communities. Participants included data \& BI team managers (five), consultants (five), analysts (two), and professors (two; one participant had a dual professor/manager role). They reported an average of 11 years of experience in data-related fields ($\mu=11.69$, $\sigma=5.66$, $min=4.5$, $max=25$).

Each session lasted approximately one hour and followed a semi-structured format with two segments. In the first segment, we introduced participants to the concept of reading order and dashboard flow, illustrated by a set of three dashboard examples featuring diverse layouts and perceived order ambiguity: (1) a long-form infographic dashboard with a clear left-right, top-down narrative layout~\cite{formative_dashboard_1}~{\small\href{https://public.tableau.com/app/profile/danielle6303/viz/Test_3050/PerformanceDash}{\faLink{}}}; (2) a classic analytic dashboard, featuring a 4x4 view grid and a row of summary metrics, including bright colored gauge views on the right~\cite{formative_dashboard_2}~{\small\href{https://community.fabric.microsoft.com/t5/Data-Stories-Gallery/COVID-19-Information-Dashboard/td-p/1405743}{\faLink{}}}; and (3) a complex, details-first infographic featuring dense data on top, and title plus context at the bottom~\cite{formative_dashboard_3} {\small\href{https://public.tableau.com/app/profile/fuadahmed/viz/TheCostofLiving/Dashboard1}{\faLink{}}}. We \rr{also} asked how reading order manifests in their own dashboard designs\rr{, including what aspects they take into account, what constitutes a ``well designed'' reading order, and the extent they explicitly consider it in their designs}. In the second segment, we \rr{showed a \cre{preliminary} visual representation of dashboard reading order as a nested graph, and asked participants to envision a similar structure to traverse components from a dashboard of their choice. Rather than collecting reading flows, our goal was to further discussion of reading order considerations} in concrete, situated scenarios. \rr{The study protocol is provided in Supplemental Materials.}

Sessions were audio- and screen-recorded and transcribed. Our transcript analysis approach first organized responses by participant into two broad categories: reflections on reading order design factors\cre{,} and reflections on authoring workflows. We then conducted a bottom-up thematic analysis within each category.

\subsection{Characterizing Reading Order \textit{Flows}}
\label{sec:interviews-design}
Participants consistently described reading order as an important\cre{,} and often deliberate\cre{,} consideration for dashboard design. They reported relying on familiar conventions aligned with visual attention~\cite{nielsen2006fpattern,itti:1998,bylinski:eyefixation:2017} and dashboard design~\cite{wexler2017big}. Most notably, 10 out of 13 participants also stated generally adhering to the notion of ``natural'' reading order~\cite{lidwell2010universal}, typically described in Western contexts as a top-down, left-to-right scan. This convention provides a coarse structural scaffold for organizing content, e.g.:%, placing high-priority information near the top of the dashboard: 
\qte{generally speaking, you will put the most important stuff on the top half and the less important stuff on the bottom half}{P9}.

While this notion may approximate average user behavior, predicting actual \textit{reading flows}, i.e., the specific orders in which users might navigate elements of a particular dashboard, is a far more challenging exercise. As one participant noted, 
\qte{here's how you read print. But then how do you read spatial elements? (...) where do you look? Because it's unguided. So pocketed, that people don't necessarily know exactly what part to look at}{P4}. Mentions of scanning patterns (e.g., F- or Z-patterns) did arise (5/13), but were mostly framed as design heuristics rather than reliable predictors of user behavior.

Individual variability was reflected in our study: participants had diverging opinions on the ``ideal'' reading order for the same dashboards. They also diverged in their design practices and in how intentionally they designed for flow; some described carefully orchestrating user interpretation, while others focused more on emphasizing key elements rather than prescribing sequence, ranging from \qte{I did definitely put a lot of thought into how someone would interpret this}{P13} to \qte{I'd put something higher, think more about the other areas to elevate, and give it more space than I would necessarily think about the order.}{P9}.

Together, these findings suggest that characterizing reading order for dashboards is a complex and multifaceted issue, and that reading order is neither fixed nor entirely designer-controlled. Instead, it emerges from an interplay between design cues and user interpretation. Importantly, this does not imply that all reading paths are equally likely. Designers actively shape user behavior by leveraging design features to nudge attention and guide transitions, while still providing room for exploration, e.g.: \qte{I made the title as prominent as possible. And then after that, I don't know if I really cared [about reading order]}{P7}. 

We report emerging areas of consensus around how practitioners influence reading flows, organized into relevant dashboard \textit{design factors} (\S~\ref{sec:formative-design-factors}) and \textit{design principles} for guiding reading order flows (\S\ref{sec:formative-design-principles}).

\subsubsection{Design Factors}
\label{sec:formative-design-factors}
One participant aptly described reading order flows as a \textit{``wayfinding''} process (\pid{P4}), where users navigate between elements based on available visual and semantic cues. This perspective aligns with prior work on wayfinding in information spaces, which characterizes navigation as a process of moving through structured environments using cues that signal possible paths and destinations~\cite{Foltz1998DesigningNI,benyon:2001}. 

We conceptualize dashboard reading order as a form of graph traversal, where dashboard elements act as nodes and transitions between them are shaped by multiple, overlapping cues. Rather than a single prescribed path, users navigate a network of possible routes, dynamically selecting where to go next based on what is perceptually prominent, semantically meaningful, or relevant to their task. In this framing, we identify six design factors that can be understood as distinct but interrelated mechanisms that shape the connections between elements and guide movement across the dashboard. They are:

\pheading{\thlayout{\textbf{Layout}}}: the structural arrangement of dashboard elements, including their relative placement, alignment, and grouping. 

\pheading{\thsaliency{\textbf{Visual Saliency}}}: the visual prominence of dashboard elements (e.g., size, color, spacing) and the extent to which they attract user attention.

\pheading{\thsemantic{\textbf{Semantics}}}: the meaning conveyed by dashboard elements and the relationships between them that support interpretation (e.g., a title establishing context for a set of charts, or a KPI summarizing \cre{associated} trends).  

\pheading{\thfunction{\textbf{Function}}}: the role of dashboard elements (e.g., titles, charts, filters, text) and how users interpret their purpose within the interface.  

\pheading{\thinteract{\textbf{Interaction}}}: user actions (e.g., filtering, hovering, drilling down) that reveal additional information (tooltips and info panels) and transform views (filtering, slicing, transforming). 

\pheading{\thcontext{\textbf{User Context}}}: external factors that influence reading order, such as context of use (e.g., first time versus recurrent use), analytical goals (e.g., specific questions versus open-ended exploration), and user characteristics (e.g., visual literacy, prior experience, personal preferences). 

%\rr{From here on, we signal references to the design themes by their corresponding mnemonics: \slayout\,\ssaliency\,\ssemantics\,\sfunction\,\sinteract\,\scontext. }

\subsubsection{Design Principles}
\label{sec:formative-design-principles}
Practitioners consistently reported placing \qte{the most important stuff}{P9} (including summary-level elements and main data views) near the top or top-left (10/13). Similarly, component proximity was often used to signal relatedness (5/13), with components placed near or next to each other to support a natural question–answer progression during sequential viewing, \qte{so that the next place that my eye is naturally drawn to answers the next question that I would have}{P5}. This aligns with established design principles such as the Gutenberg Diagram~\cite{lidwell2010universal} which identifies high-attention regions, and also Shneiderman's ``overview first" mantra \cite{shneiderman2003eyes}, which emphasizes prioritizing summary information early in the viewing process.

\pheading{\thfunction{\textbf{Dashboard titles}} serve as \thlayout{\textbf{structural}} and \thsemantic{\textbf{semantic}} ``anchors.''} Practitioners emphasized the importance of designing titles to be visually prominent and consistently placed to support a smooth reading flow. Many (5/13) considered titles to be the primary entry point to dashboards, fulfilling two important roles. First, they provide structural anchoring, establishing a clear starting point for navigation: \qte{I'm gonna capture you with the title, at the right the way top, at the bar at the top}{P2}. Second, they provide semantic orientation, helping users quickly understand the topic: \qte{I find myself looking for the title, because I want to know what this is about}{P7}.

\pheading{\thsaliency{\textbf{Visual saliency}} modulates attention beyond \thlayout{\textbf{layout}}.} While layout establishes structural expectations, visual saliency allows designers to override or refine these expectations. Most practitioners (9/13) reported making use of visual design features as a way to
\qte{give some sort of priority or dominance based on your color, your size, or the flow}{P6}.
When combined with layout, saliency becomes a powerful mechanism for guiding attention. However, visual saliency can also compete with layout cues, and conflicts between them lead to ambiguity or confusion. 
As an example, several participants (4/13) commented on the attention split caused by the gauge saturation in one of the dashboard examples used~\cite{formative_dashboard_2}, e.g., \qte{even if the tendency is to start from the left corner, my eye is drawn to the brightness of those gauges}{P7}.

\pheading{\textbf{Know your \thcontext{\textbf{audience}}.}} 
Understanding the audience was one of the most frequently cited considerations (9/13). 
Practitioners mentioned tailoring reading order based on \thcontext{user goals, domain familiarity, and cultural conventions}. A common concern was understanding analytical goals (4/13), which directly informs how \thsemantic{content}, or the \qte{logical flow for analysis}{P12} is placed: \qte{There's a question that brings you to this. And I want the answer to my immediate question to be right there, at the top, or top left}{P5}. 
Others report building on familiarity to guide their designs, to shape future expectations:  
\qte{We've created dashboard starter templates [where] the controls are all in the same place, because we want to create muscle memory}{P9}.
 
\pheading{\thinteract{\textbf{Interaction design}} for a \thcontext{\textbf{seamless user experience}}}. 
Dashboard interaction becomes essential 
to navigate \rr{large amounts of data and let} \qte{people choose which view they want to see}{P2}. \rr{But its influence on reading order was sparsely mentioned, and the few instances gathered offered little consensus: e.g.,} filter placement preferences ranging from the \textit{top} (\pid{P3}), the \textit{left} (\pid{P13, P2}), the \textit{right} (\pid{P1, P2}), or \textit{either} (\pid{P5}).
We posit that the branching nature of interactivity makes it difficult to predict and plan for in the context of reading order, and calls for a need to approach it holistically to prioritize user experience and match expectations. To that, participants brought up notions of \thcontext{familiarity}, with users exploring data to \qte{get to [relatable] data they know}{P4}; \qte{rewarding curiosity}{P9} by providing details on demand; and avoiding ``quick'' filters (i.e., side filtering widgets) to not \qte{take the client out of the flow of analysis}{P11}. 

%% file: tex/04-author-user-studies.tex
\begin{figure}
    \centering
    \includegraphics[width=\linewidth]{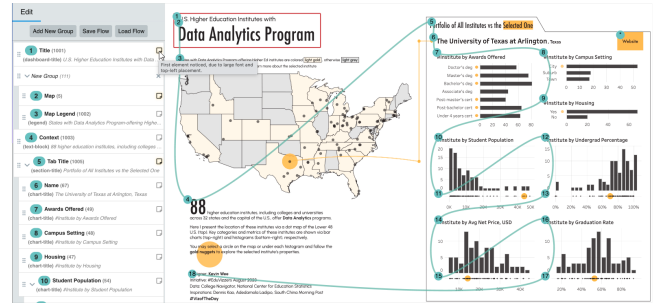}
    \caption{Custom flow specification tool used in the data collection studies.}
    \label{fig:validation-collection}
    \vspace{-1em}
\end{figure}

\begin{figure*}
    \centering
    \includegraphics[width=\linewidth]{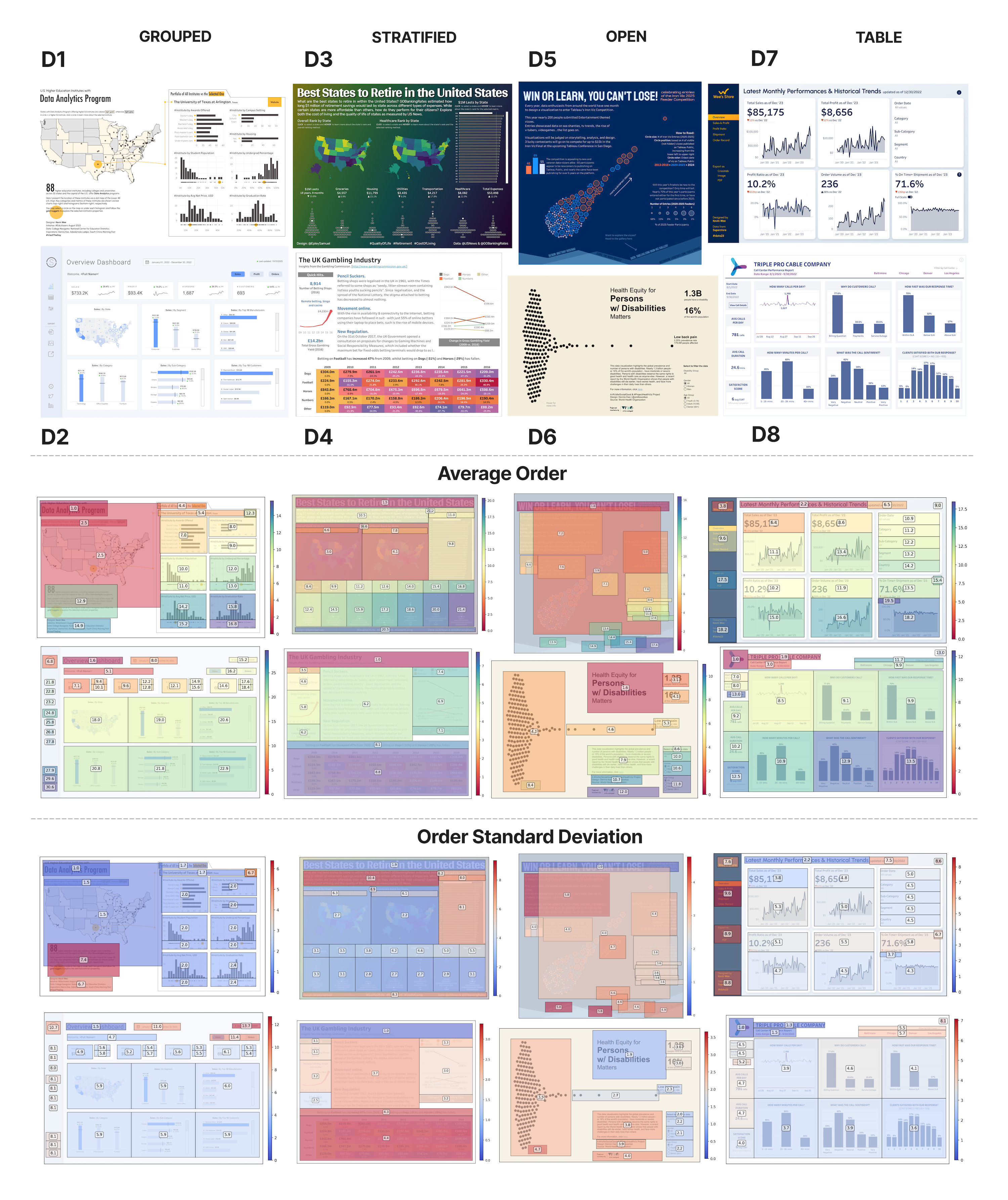}
    \caption{\textbf{Top row}: Dashboards used in the study (\textbf{D1}-\textbf{D8}). Zoom in for details. 
    \\   \textbf{Middle row}: Average order maps for dashboard components, spanning {\color[HTML]{D53E4F}\textbf{first}} 
    \includegraphics[width=10em, height=.75em]{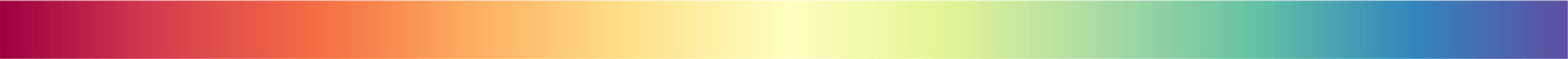}
    {\color[HTML]{5E4FA2}\textbf{last}} in order.
    %{\color[HTML]{DA5E57}\textbf{red}} (lower order~=~visited \textit{earlier}) to {\color[HTML]{4069D3}\textbf{blue}} (higher order~=~visited \textit{later}). 
    \\
    \textbf{Bottom row}: Standard deviation maps for dashboard component orders; spanning {\color[HTML]{3b4cc0}\textbf{lower}}
    \includegraphics[width=10em, height=.75em]{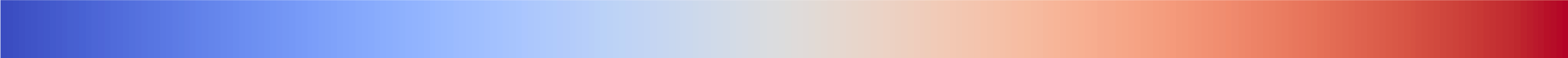}
    {\color[HTML]{b40426}\textbf{higher}} variance.
    %transitions from {\color[HTML]{4069D3}\textbf{blue}} (lower variance) to {\color[HTML]{DA5E57}\textbf{red}} (higher variance). 
\\    
\textit{\textbf{Dashboard Credits:}\\}
    \textbf{D1}: U.S. Institutes with Data Analytics Program
    \href{https://public.tableau.com/app/profile/kevin.wee/viz/USHigherEdInstituteswithDataAnalyticsProgram/Dashboard}{[link]} by
    \href{https://public.tableau.com/app/profile/kevin.wee}{Kevin Wee}.
    \textbf{D2}: Superstore Overview
    \href{https://public.tableau.com/views/SuperstoreDashboard_16709573699130/SuperstoreDashboard}{[link]} by
    \href{https://public.tableau.com/app/profile/p.padham}{Priya Padham}.
   \\ %
    \textbf{D3}: Best States to Retire in the United States
    \href{https://public.tableau.com/app/profile/sam.epley/viz/BestStatestoRetireintheUnitedStates/BestStatestoRetireintheUnitedStates}{[link]} by
    \href{https://public.tableau.com/app/profile/sam.epley}{Sam Epley}.
    \textbf{D4}: The UK Gambling Industry
    \href{https://public.tableau.com/app/profile/scribblr.42/viz/TheUKGamblingIndustry/TheUKGamblingIndustry}{[link]} by
    \href{https://public.tableau.com/app/profile/scribblr.42}{Ravi Mistry}.
   \\
        \textbf{D5}: Win or Learn, you Can't Lose!
    \href{https://public.tableau.com/app/profile/brrosenau/viz/Celebratingthe2025IronVizFeederEntries/CelebratingIronViz2025Participants2}{[link]} by
    \href{https://public.tableau.com/app/profile/brrosenau}{Brittany Rosenau}.
    \textbf{D6}: Health Equity for Persons w/Disabilities Matters.
    \href{https://public.tableau.com/app/profile/dennis.kao/viz/HealthEquityforPersonswDisabilitiesMatters/Dashboard1}{[link]} by
    \href{https://public.tableau.com/app/profile/dennis.kao}{Dennis Kao}.
    \\
    \textbf{D7}: Wee's Store Latest Monthly Performance and Historical Trends
    \href{https://public.tableau.com/app/profile/kevin.wee/viz/Data23DemoCreativeUsesofImagesinaTableauDashboard_16828225482900/Overview}{[link]} by
    \href{https://public.tableau.com/app/profile/kevin.wee}{Kevin Wee}.
    \textbf{D8}: Triple Pro Cable Company
    \href{https://public.tableau.com/app/profile/sarah.pallett/viz/RWFDCallCenter_16797900072550/RWFDCallCenterDashboard}{[link]} by
    \href{https://public.tableau.com/app/profile/sarah.pallett}{Sarah Pallett}.
    \\
 }
    \label{fig:validation-dashboards}
\end{figure*}

\section{Capturing Dashboard Reading Order}
\label{sec:validation}

Formative findings \cre{identified design} factors and principles that influence dashboard reading order and can inform authoring practices for guiding user attention and flow. 
\rr{Eye-tracking studies underscore the effect of 
\thlayout{layout}~\cite{zhang2024effects} and \thsaliency{saliency}~\cite{bylinski:eyefixation:2017, yang2025dashboard} on scanning behaviors, including fixation density over time for different dashboard objects which suggests a role of \thfunction{function} in scanning patterns~\cite{yang2025dashboard}. These factors were looked at statically or in isolation, providing an incomplete picture of how design patterns influence each other and how users \textit{reason} over these factors as they navigate through dashboards.}
%these practices align with actual \textit{end-user reading flows}.} 

To \rr{inform these gaps, we opted for a qualitative, think-aloud} data collection study. For a diverse set of eight dashboards (\S\ref{sec:validation-dashboards}), we developed an annotation tool (\S\ref{sec:validation-tool}) to collect \textit{reading flow instances} from both dashboard authors (\S\ref{sec:validation-authors-methodology}) and 16 end-users who routinely work with dashboards (\S\ref{sec:validation-end-users-methodology}).
Each instance represents how an individual user traverses a dashboard, enabling us to analyze shared patterns and points of divergence across users. For brevity, we refer to these as \textit{flow instances} throughout the remainder of the paper.

\subsection{Sourcing the Dashboards}
\label{sec:validation-dashboards}
Eight dashboards (\db{1}-\db{8}, \autoref{fig:validation-dashboards}) were sourced from seven authors (\pid{A1}-\pid{A8}: A1 is also A7). To ensure access to authors (for their reading order design rationales), we first identified eligible volunteers through multiple recruitment channels, and then selected dashboards from their publicly available work on Tableau Public~\cite{tableau_public}. Recruitment channels included (a) a call for volunteers to self-nominate their own dashboards, posted in practitioner communities (three authors); (b) re-recruiting participants from the formative study (two authors); and (c) directly reaching out to practitioners in our network (two authors).
 
To ensure diversity, we curated dashboards spanning Bach et al.'s layout typology~\cite{bach2022dashboard}, including \textit{grouped} (\db{1}, \db{2}), \textit{stratified} (\db{3}, \db{4}), \textit{open} (\db{5}, \db{6}), and \textit{table} (\db{7}, \db{8}) layouts, as well as multiple genres (from analytic to infographic), domains, chart types, and design styles. We intentionally chose dashboards exhibiting reading order ambiguity through non-standard layouts (e.g., title placement on the right \db{6}, unusual chart–text alignments in \db{3} and \db{4}, KPI boxes on the left in \db{8}) and flow-altering features (e.g., enclosures in \db{1}, annotations in \db{5}, sidebars in \db{2} and \db{7}).

\subsection{Annotation Tool}
\label{sec:validation-tool}

We created a custom flow specification tool \rr{used by study moderators} to collect reading flow instances from authors and end-users.
It displays a dashboard alongside a re-orderable list of dashboard components, and overlays the corresponding order as a numbered spline to represent the reading flow path (\autoref{fig:validation-collection}). We used the Tableau Embeddings API~\cite{tableau_embedding_api} to retrieve the JSON list of dashboard zones as an initial components list. We then manually augmented this list to include dashboard elements not originally captured as zones (e.g., text embedded in background images). These components lists were subsequently validated by dashboard authors to ensure completeness and appropriate granularity. Only one dashboard required refinement (\db{6}), where the author suggested splitting the beeswarm chart into two sub-components to distinguish the dense cluster on the left from the outliers on the right.

Flow instances are specified by \textit{reordering} the list of dashboard components, so elements \rr{``visited''} early are placed further up on the list. The tool additionally provides features to \textit{skip} components (i.e., remove them from flow), which we used when dashboard elements were missed or ignored; and optional features to \textit{nest} and \textit{group} components. Because these hierarchical features were used sparingly by end-users, we excluded them from our flow instance analysis.  

In both author and end-user sessions, the moderator operated the tool via screen sharing, with participants verbally indicating their reading order. This ensured consistency while reducing participant burden, and allowed us to pause on interactions (e.g., clicks, hovers) to probe reasoning that might otherwise remain implicit. 

\subsection{Author Sessions}
\label{sec:validation-authors-methodology}
We invited the seven dashboard authors to participate in 45-minute remote sessions to capture their \textit{intended} reading order flows and underlying design rationales. Our goal with these sessions was to understand how design intent aligns with or diverges from end-user reading flows. Participants included analysts and engineers (four), as well as data-related positions in design (one), management (one), and education (one). They reported an average of approximately 10 years of experience in data-related fields (min=5, max=15).

Authors were asked to \rr{specify} one or more flow instances representing the reading flows intended in their designs, while thinking aloud and explaining the reasoning behind design decisions. \rr{Most authors chose to provide two alternative flows, each for a different use case; more details in \autoref{fig:appendix-author-flows-1} and \autoref{fig:appendix-author-flows-2} in the Appendix.} These sessions provide insight into how dashboard authors conceptualize and attempt to guide reading order through layout, content, and interaction. \rr{The study protocol is provided in Supplemental Materials.}

\subsection{End-User Sessions}
\label{sec:validation-end-users-methodology}
Following the author sessions, we recruited 16 participants (\pid{E1}–\pid{E16}) from a large B2B technology company (> 50,000 employees) who are regularly \textit{``exposed to data, charts, and dashboards as part of their work''} for one-hour remote sessions. Participants spanned roles including analysts, developers, and managers, and self-identified as primarily dashboard \textit{consumers} (four), \textit{creators} (six), or \textit{both} (six).

Each participant was assigned four out of the eight dashboards. Dashboards were counterbalanced so every participant was exposed to at least one dashboard from each of the four layouts (\S\ref{sec:validation-dashboards}), resulting in 16 unique dashboard combinations (i.e., one per participant) and yielding a total of \textbf{8} \textit{flow instances} per dashboard (i.e., an overall total of \textbf{64} flow instances). Dashboard presentation order was randomized.

Participants were asked to report their reading-order flows for each dashboard, which were captured by the moderator using the flow specification tool. We framed the task \rr{as a ``first contact'' with a dashboard, instructing them to think} about \textit{``the natural way you would go through this dashboard to understand it''}, and to specify \textit{``the order you would go through these dashboard components, in a way that is logical and makes sense to you''}. Before each segment, participants were told to \textit{``take a moment to see what this dashboard is about''} and were encouraged to think aloud. \rr{When ready, participants narrated a sequence starting with where to begin the flow, then where to go next, and why. The moderator located and ordered the respective elements. When finished, the moderator validated the captured order, and listed remaining elements to either order or flag them as ``skipped" (e.g., if element was not noticed or purposefully ignored).}
%If necessary, participants were asked to provide further rationale for their choices
Each segment lasted approximately 10 minutes. \rr{Segments were preceded by a training session, and were followed by a discussion} on the flow authoring experience, including heuristics used and any challenging ordering instances. \rr{The study protocol is provided in Supplemental Materials.}

\subsection{Data Analysis}
%Each flow instance was represented as a flat, ordered sequence of dashboard components, where each component appears at most once. In cases of circular flows (i.e., when participants \rr{mentioned} a component before completing the flow\rr{, for example when revisiting a filter after an overview of its associated data views}), we recorded only the first occurrence of each component. This simplification allowed us to standardize representations across participants and focus on primary traversal paths. We compared individual flow instances for the same dashboard to surface patterns. We also examined aggregate statistics, including the mean and standard deviation of component order indices across instances, revealing recurring patterns and points of divergence (\autoref{fig:validation-dashboards}). To complement our flow instance analysis, we extracted rationales from transcripts. We summarized these into structured notes consisting of ordered elements and associated rationales, and then selectively coded them according to the six reading order design factors derived from the formative study (\S\ref{sec:formative-design-factors}). This enabled us to identify per-participant patterns and provide contextual grounding for the observed flow structures. It also supported comparisons between author-intended and end-user reading flows.
Each flow instance was represented as a flat, ordered sequence of dashboard components, where each component appears at most once. In cases of circular flows (i.e., when participants \rr{``revisited"} a component before completing the flow; \rr{e.g., wanting to return and interact with a filter after a first overview pass}), we recorded only the first occurrence of each component. This simplification allowed us to standardize representations across participants and focus on primary traversal paths. We compared individual flow instances for the same dashboard to surface patterns. We also examined aggregate statistics, including the mean and standard deviation of component order indices across instances, revealing recurring patterns and points of divergence (\autoref{fig:validation-dashboards}). To complement our flow instance analysis, we extracted rationales from transcripts. We summarized them into structured notes consisting of ordered elements and associated rationales, and then selectively coded them according to the six reading order design factors from the formative study (\S\ref{sec:formative-design-factors}). This allowed us to identify per-participant patterns and explain choices for observed flow structures. It also supported comparisons between author-intended and end-user reading flows.

%% file: tex/05-findings.tex
\section{Characterizing Reading Order Flow Patterns}
\label{sec:flow-findings}
We synthesize observed reading behaviors into a set of recurring flow patterns that characterize how users traverse dashboards. Rather than following a single linear path, participants exhibited structured yet flexible navigation strategies shaped by the interplay of layout, semantic cues, visual prominence, \rr{interaction affordances}, and task context. \rr{Across dashboards, users consistently converged on common entry points and local groupings, while diverging primarily in later transitions where multiple plausible analytical paths existed.} %These patterns reveal strong commonalities such as consistent entry points and local groupings, and meaningful divergences in how users transition between components.

Reading order \rr{is therefore best understood as a set of \emph{probabilistic pathways} rather than a single canonical sequence}. While certain elements reliably anchor the start of exploration (e.g., titles, KPIs), subsequent transitions depend on how users interpret relationships between views, prioritize information, and pursue their goals. This perspective aligns with prior models of iterative analysis~\cite{pirolli2005sensemaking} and work on coordinated multiple views, which emphasize the role of spatial and semantic relationships in guiding navigation~\cite{roberts:2007}.

Across dashboards, layout provides an initial scaffold for traversal but does not fully determine reading order. Instead, users combine \thlayout{structural cues} (e.g., top-left positioning), \thsemantic{semantic signals} (e.g., titles and labels), \thsaliency{visual features} (e.g., size, color, or chart type), \rr{and \thinteract{interaction}} to construct their own reading paths. This results in flows that are locally coherent where related elements are \rr{seen together, yet globally variable} in regions where multiple plausible next steps exist.

\rr{The following sections characterize these recurring behavioral patterns and underlying mechanisms, in the context of a ``first dashboard contact'' task. Rather than propose design rules, we explain how interactions among \thlayout{layout}, \thsemantic{semantics}, \thsaliency{visual saliency}, \thfunction{function}, \thinteract{interaction}, and \thcontext{user context} shape reading flow, providing a behavioral foundation for understanding, evaluating, and designing dashboard navigation.} %In the following sections, we unpack these patterns in detail, highlighting where reading behaviors converge and diverge. In doing so, we move beyond aggregate summaries to surface the underlying mechanisms that shape dashboard traversal and revisit existing design principles in light of these findings.

\subsection{Emerging Flow Patterns and Design Implications}
Our mixed-methods analysis of flow instances revealed both recurring patterns in how users traverse dashboards and areas of meaningful divergence. By examining ordering statistics alongside individual flow instances and associated rationales, we \rr{first examine dominant patterns shared across participants (\S\ref{sec:flow-findings-dominant}), followed by recurring variants that reflect differences in user interpretation and analytical goals (\S\ref{sec:flow-findings-variants})}. %revisit the design principles introduced in \S\ref{sec:formative-design-principles} with more nuance (\S\ref{sec:flow-findings-dominant}) and characterize how distinct flow strategies manifest (\S\ref{sec:flow-findings-variants}).

\subsubsection{Dominant Patterns}
\label{sec:flow-findings-dominant}
\rr{We begin with dominant reading flow patterns observed across participants,
%. Rather than prescribing new design rules, these findings 
characterized as recurring behavioral mechanisms that consistently shaped early and mid-stage traversal. }\cre{Notably, these findings revisit, and further substantiate, the design principles in \S\ref{sec:formative-design-principles}.}
%Later sections examine where these patterns diverge and how user goals, context, and dashboard design give rise to alternative reading strategies.} %focusing on areas of strong agreement in reading order flows. These patterns highlight consistent strategies that users rely on to initiate and structure their traversal, revealing how layout, semantics, and visual cues jointly shape early and mid-stage navigation. While later sections examine sources of variability, the patterns presented here reflect stable behaviors that can serve as reliable design foundations.

\pheading{\textbf{Layout-driven progression} as a dominant heuristic.}
Order averages (\autoref{fig:validation-dashboards}-middle) indicate a strong influence of \thlayout{``natural'' reading order}, with a general left-to-right, top-to-bottom progression. This pattern is especially pronounced in business dashboards where layout aligns with \thsemantic{information granularity}, for example by placing summary-level elements (e.g., KPIs, BANs) near the top (\db{2} and \db{7}). These findings reinforce the design principle identified in \S\ref{sec:formative-design-principles}.

This tendency is also observed, albeit less strongly, in more open layouts (\db{5} and \db{6}), where the alignment between layout and semantic structure is less explicit. In such cases, participants appear to fall back on natural reading order as a default scaffold for traversal, particularly when the dashboard structure or content is less familiar.
% \implication{1}{\thlayout{Top-down} $\leftrightarrow$ \thsemantic{overview to details} alignment is a strong design default for minimizing reading friction. Organizing dashboards to present summaries (e.g., KPIs or overview charts) before progressively more detailed views supports a reliable and intuitive flow, particularly for general audiences and first-time use.}
\rr{\implication{1}{Reading flow is most predictable when \thlayout{spatial organization} reinforces \thsemantic{semantic hierarchy}. Designers seeking to encourage consistent overview-to-detail navigation should align layout with semantic organization, particularly for first-time users.}}

\pheading{Consistent entry points aligned with layout and overview.}
\thfunction{Titles}, \thfunction{headers}, and primary \thsemantic{overview} components were consistently visited first, as evidenced by average and standard deviation order maps (\autoref{fig:validation-dashboards}-bottom).  Most flows began with either the {title} (42/64, 65.6\%) 
or with adjacent {logo $\rightarrow$ title} sequences (17/64, 26.6\%).
Exceptions to this pattern emerged in \db{3} (2/8) and \db{6} (3/8), and were all \thsaliency{saliency} + \thlayout{layout} related, with end-users choosing to start glancing over visually salient charts near the top or top-left before moving on to the title. For instance, for \db{3}: \qte{The first thing that my eyes are drawn to are the two maps. But wouldn't go into too much detail (...) and then I would see the title.}{E16}.
Regarding \db{6}, it is worth noting that, despite not being on the preferred top-left location, the title still figures as a starting point in most flow instances (5/8). Many end-users (7/16) also explicitly mentioned starting from the title as a personal reading order strategy, \cre{e.g.,}
\qte{so I have at least a general idea of what the dashboard is about}{E1}. These findings confirm dashboard titles and associated header-level elements as reliable spatial and semantic anchors to dashboard understanding. 
% \implication{2}{\thfunction{Titles, logos, and subheadings} consistently act as anchors in reading flows. Designers should treat these elements as primary guidance mechanisms, ensuring they are both prominent and semantically meaningful, by carefully considering the \thlayout{placement}, \thsaliency{saliency}, and \thsemantic{framing} of these elements through concise, high-signal content.}
\rr{\implication{2}{\thfunction{Titles, logos, and other header-level elements} consistently establish the entry point into dashboard reading flows. They function as \thsemantic{semantic anchors}, helping users orient themselves before exploring details. Designers seeking to support rapid orientation should reinforce this anchoring through the alignment of \thlayout{placement}, \thsaliency{visual saliency}, and concise \thsemantic{semantic framing}.}}

\pheading{Visual, semantic, and spatial proximity drive grouping consistency}. \rr{Participants consistently interpreted \thlayout{nearby} elements that were \thsaliency{visually} and \thsemantic{semantically} similar as coherent groups.} This pattern was evident across chart grids (\db{1}, \db{2}, \db{8}), filter blocks (\db{6}, \db{7}), and the upper left and right columns of components in \db{4}. \rr{Within these groups, participants also exhibited highly consistent local reading orders, typically following a left-to-right, top-to-bottom progression.}
However, less agreement between the three design factors also mapped to less consistency in ordering behaviors; we discuss emerging variants in \S\ref{sec:flow-findings-variants}. \rr{Together, these findings indicate that participants interpreted dashboards as collections of related information groups rather than independent visual elements.} %Overall, this pattern reflects a tendency to interpret dashboards not as collections of independent elements, but as structured groupings of related information. 
% \implication{3}{Aligning \thsaliency{consistent visual design} with intentional \thlayout{spatial proximity} may increase the likelihood that \thsemantic{logically related elements} will be seen and interpreted together.}
\rr{\implication{3} {Reading flow is most consistent when \thlayout{spatial proximity}, \thsaliency{visual similarity}, and \thsemantic{semantic relatedness} reinforce one another. Designers seeking to encourage coherent local reading flows should align these cues so that related elements are more naturally perceived and interpreted as a group.}}

\pheading{Consistent anchoring on salient \textit{``big aggregate numbers''} (BANs).}
 \thfunction{BANs} are \rr{a common mechanism for conveying aggregate metrics and }\thsemantic{overview-level content} in business dashboards. \rr{Across our corpus, these} components exhibited low average order and \rr{order} variance, indicating a general tendency to prioritize them in the reading order. \thsaliency{Visually} and \thlayout{spatially} prominent BANs (e.g., in \db{2} and \db{6}) were consistently visited early, shortly after the title, suggesting they are favored in both business and infographic dashboards alike. In \db{8}, there is seemingly weaker consensus in flow instances for less visually and spatially prominent BANs such as the KPI boxes which came \textit{after} charts half the time (4/8); however, participant feedback points to these elements being initially missed in the flow rather than deliberately de-prioritized, due to the location, e.g., KPIs \qte{on top of the dashboard}{E15}; or lower saliency, e.g., \qte{there's nothing distinguishing them color- or size-wise from the larger charts}{E3}. \rr{These observations suggest that the role of BANs as flow anchors depends on the alignment of their semantic importance with their visual prominence and placement.}
% \implication{4}{\thfunction{BANs} are useful to convey \thsemantic{overview-level information} to end-users. When present in dashboards, they should be carefully designed to either \thsaliency{visually stand out}, be \thlayout{prominently placed}, or (ideally) both.}
\rr{\implication{4}{ When \thfunction{BANs} are intended to establish an overview of the dashboard, reading flows are most consistent when their \thsemantic{semantic} and \thfunction{functional}  importance is reinforced through \thsaliency{visual prominence} and \thlayout{spatial placement}. Designers seeking to support rapid orientation should align these cues so that overview metrics naturally function as early anchors.}}

%\pheading{\slayout\,\ssaliency\,\ssemantics\,\sfunction\,Consistent anchoring on salient \textit{``big aggregate numbers''} (BANs).}
% BANs are prevalent dashboard features to convey aggregate metrics, and the most common form of  {overview-level content} in business dashboards. 
%These components exhibited low average order values and low variance, indicating a general tendency to prioritize them in the reading order. Visually and spatially prominent BANs (e.g., in \db{2} and \db{6}) were consistently visited early, shortly after the title, suggesting they are favored in both business and infographic dashboards alike. In \db{8}, there is seemingly weaker consensus in flow instances for less visually and spatially prominent BANs such as the KPI boxes which came \textit{after} charts half the time (4/8); however, participant feedback points to these elements being initially missed in the flow rather than deliberately de-prioritized, due to the location, e.g., KPIs \qte{on top of the dashboard}{E15}; or lower saliency, e.g., \qte{there's nothing distinguishing them color- or size-wise from the larger charts}{E3}.
%\slayout\,\ssaliency\,\ssemantics\,\sfunction\, \implication{4}{{BANs} are useful to convey {overview-level information} to end-users. When present in dashboards, they should be carefully designed to either {visually stand out}, be {prominently placed}, or (ideally) both. \slayout\,\ssaliency\,\ssemantics\,\sfunction\,}

\subsubsection{Reading Order Variants}
\label{sec:flow-findings-variants}

While participants exhibit strong agreement on certain flow decisions, such as using layout as a scaffold and titles as entry points, \rr{reading flows diverged in later stages of exploration}. In fact, no two flows look exactly the same, particularly among the more unorthodox dashboards (e.g., \db{5} and \db{6}, \autoref{fig:appendix-db-user-flows-2}-\textit{Appendix}). %But a deeper look at flow subsequences reveals a few clusters of partial agreement, further qualified by end-user rationales. We found mid-flow choices to be largely influenced by \thcontext{personal preference}, modulating the degree that \thsaliency{gestalt patterns} and \thsemantic{information granularity} affect flows at an individual level.
\rr{Examining common flow subsequences alongside participant rationales revealed that mid-flow decisions are strongly shaped by \thcontext{personal preferences},
modulating the degree that \thsaliency{Gestalt patterns} and \thsemantic{information hierarchy} affected individual flow strategies. 
Oftentimes, these decisions diverged across participants in  predictable ways as they pursued overlapping goals and styles. %were largely stable across different flows for the same individual, 
% These findings indicate that dashboard reading behavior is characterized by structured variability. 
%users share common semantic anchors and local traversal patterns, 
} %These preferences were generally stable across dashboards for the same user, which calls for a better understanding of these variants of partial consensus. 

\pheading{Prominent visuals attract attention and can override layout order}.
Components with \thsaliency{high visual prominence} tended to attract attention early. The large maps in \db{3} and the beeswarm chart in \db{5} were often the first elements visited after the title, overriding other textual and visual elements despite \rr{the \thlayout{less prominent placement}}. In these cases, participants consistently reported deviating from linear scanning to first attend to elements perceived as more informative or attention-grabbing.

\thfunction{Annotations and connectors} also had an influence, e.g., the bottom annotation on \db{5} pointing to source metadata redirected flow earlier to that metadata piece half of the time (4/8).
This effect was made explicit for the \thsaliency{yellow link} in \db{1}; instead of following the layout top-down, most flows (6/8) transitioned to the right side panel, with half (3/8) attributed to the yellow connector in participant rationales. 

There are confounds to these findings worth discussing. First, the effect appears to be \thlayout{adjacency-scoped}, which is evidenced by the \rr{very salient yet} bottom-located heatmap in \db{4} still consistently visited last or near-last (6/8). 
%This points to a nuance in \thcontext{personal preferences} that chart attention may be less about saliency attraction, and more about text avoidance. 
Second, saliency attention can actually be destabilizing. In 
four %E1, E3, E10, E16, ~E11, ~E14,
separate occasions, participants perusing \db{5} mentioned being initially attracted to the colorful beeswarm, only to realize they don't know what they are looking at: 
\qte{As I'm reading the title, my eye immediately goes to the little red dot \emph{[outlier, top right]}. So now I'm all discombobulated. I gotta go back \emph{[to the top left]}}{E3}.
Third, end-user rationales underscored that not all users were necessarily predisposed to visual saliency as a flow driver, choosing to instead prioritize layout, semantic continuity, or function: e.g., in \db{6}, most end-users (5/8) still chose to start their flow from the \thfunction{title}, and not the salient beeswarm on the left. %\rr{Interestingly, many flow deviations in \db{1}, \db{3}, and \db{5} were justified not as a visual saliency effect but as \thfunction{text avoidance}, e.g., in \db{5}: \qte{I'm not really a big text person, so I'd probably jump down to that bubble chart}{E1}.}
%as a desire to bypass \thfunction{large blocks of text} rather than 
These observations underscore the influencing power (and limits) of visual saliency as a versatile tool to signal alternative flows, and a need to leverage it responsibly.
% \implication{5}{\thsaliency{Visual saliency} through color, size, and style is a useful tool to direct local attention and signal secondary flows. As such, it should be used with caution, as it can also disorient users.}
\rr{\implication{5}{\thsaliency{Visual saliency} influences reading flow most effectively when it reinforces \thsemantic{semantic} context rather than competing with it. Salient elements can successfully redirect \thlayout{local attention} and highlight alternative analytical paths, but excessive or context-free saliency may disrupt orientation and fragment otherwise coherent reading flows.}}

%\pheading{Semantics drive low-level ordering, but strategies vary}.
\pheading{Semantics shape reading flow beyond initial orientation.} \rr{Participants generally followed Shneiderman's \thsemantic{``overview first'' principle}~\cite{shneiderman2003eyes}, beginning with titles and summary-level components such as BANs before progressing to more detailed views. While semantic cues consistently guided early reading decisions, \thcontext{individual strategies} became increasingly diverse as reading progressed.} % in their flows. This is evidenced in the way that titles and summary-level components like BANs were, on average, visited first, indicating the role that semantics play in guiding early flow decisions. However, these signals get murkier as flow progresses.
\rr{A closer look at flow instances and rationales revealed diverse preferences in the degree of analytic depth pursued, from establishing what the dashboard is about and obtaining a high-level snapshot of the data, to inspecting data, exploring interactions, vetting quality and provenance, and forming explanations by relating views. These strategies influenced how participants prioritized components throughout the dashboard. For example, overview-leaning individual\cre{s} consistently ``broke'' \thsemantic{semantic} and \thsaliency{visual} BAN+chart \cre{groupings} in \db{3} (5/8) and \db{7} (3/8) to instead prioritize all BANs first. Similarly, the ``how to read'' instructions in \db{5} were often visited before other contextual information despite not following the surrounding layout. These semantic reading strategies remained relatively consistent for individual participants across dashboards, suggesting that users develop stable preferences for how they acquire and organize information during analysis.}
 \rr{\implication{6}{Reading flow reflects stable semantic strategies in addition to \thsaliency{visual} and \thlayout{spatial organization}. Designers seeking to support \thcontext{diverse audiences} should consider how overview information, contextual explanations, and detailed analysis are organized to accommodate different \thsemantic{information-seeking strategies}. %rather than assuming a single canonical path through the dashboard.
 }}

\pheading{Short text \rr{consistently supports orientation, while long text produces polarized reading behaviors}.}
End-users were consistently drawn to short text in the form of \thfunction{titles and BANs} as primary sources of \thsemantic{summary-level information}. On the other hand, \thfunction{long text paragraphs} (present in \db{1}, \db{3}, \db{4}, \db{5}, and \db{6}) exhibit high order variance and relatively higher average order compared to their neighbors, \rr{often denoted in rationales as explicit} \textit{text avoidance}, e.g.: \qte{if the text is too large, I will tend to skip over it}{E13}. Looking at individual flow instances, we see that while not unanimous behavior, long paragraphs are often left for last, even when prominently placed near the top or left:
in D1 (5/8), %E1, E15, E3, E5, E7
D3 (3/8), % E1, E10, E6, 
and D5 (3/8). %E1, E10, E16, 
Conversely, a small but non-negligible number of end-users consistently favored long text, visiting them early in their flow more than once (3/16). Once again, both behaviors were consistent enough across flows from the same participant to consider them another facet of \thcontext{personal preferences}.
% \implication{7}{Dashboard authors should use \thfunction{short text} to \thsemantic{orient and frame interpretation} early in the flow, while placing \thfunction{longer explanatory text} in \thlayout{close proximity} to relevant visuals or supporting roles, rather than as primary entry points.}
\rr{\implication{7}{\thfunction{Short text} reliably establishes \thsemantic{orientation}, whereas \thfunction{longer explanatory text} supports \thsemantic{deeper interpretation} for a subset of users. Designers should use concise text to establish context early in the reading flow, while integrating longer explanations with the visual elements they support rather than assuming they will serve as universal entry points.}}

\pheading{\rr{Interpretation begins before interaction, but interaction planning begins early}.} \thinteract{Interactive components} such as filters, tabs, and selectors occupied intermediate positions in the reading order, reflecting their role as tools for refinement.  
Their placement in the flow varied somewhat across dashboards, but showed little variability across flows for the same dashboard, as evidenced by standard deviation maps (\autoref{fig:validation-dashboards}-bottom).  
For example, the more prominent \thfunction{filters} in \db{7} were consistently visited before trend charts (6/8), and often after BANs (4/8); whereas filters in \db{6} figured more often after the paragraph text to its left (5/8) than before, while being consistently among the last components to be visited. Conversely, %the filters icon in \db{2} was generally visited with its neighbors, and not often immediately recognized as a toggle for the filter panel; and 
the drop downs next to the map and bar chart in \db{3} were often visited last (3/8) % E4, E6, E13
or missed entirely (3/8), %E7, E10, E16
due to their small size.
Overall, these flows are generally aligned with standard expectations for \thlayout{layout} and \thsaliency{saliency}, given their placement and design.

However, we establish a distinction between how early these interaction-related components figure in the flow, and how early end-users would have \thinteract{interacted} with \textit{any} component of the dashboard  as part of their flow. Our data collection strategy precluded detailed insights into such interactivity patterns, but end-user rationales provide reasonable evidence that interactivity happens early and often. 
All participants alluded at least once to the use of interactive features mid-flow,
e.g., asking the moderator to hover over data points in \db{5} and \db{6} to better understand the chart encodings; pausing on filters and reporting how they would have interacted with them, or circling back for closer exploration after taking in other parts of the dashboard. 
We posit these interactions are closely associated with our earlier observations on \thsemantic{analytical depth} \thcontext{preferences}, and can vary widely. On the other hand, the observation that interactions can occur early has implications for \thlayout{placement} of interactives. For example, some users noted that interaction guidance was not perceived early on in \db{5} and \db{6}, as they navigated (and struggled to interpret) the unfamiliar beeswarm encoding.
% \implication{8}{Consider when and where end-users may \thinteract{interact}, and use the \thlayout{surrounding area} for \thsaliency{prominent} \thfunction{filter} placement, guidance, and other scaffolds.}
\rr{\implication{8}{Reading flow and \thinteract{interaction readiness} do not necessarily coincide. While interaction controls are often inspected later in the reading flow, users may begin planning exploratory interactions much earlier. Designers should therefore provide \thsaliency{interaction cues}, guidance, and supporting scaffolds \thlayout{near} the points where users first form interpretations, rather than relying solely on the location of \thfunction{filters or controls}.}}

\pheading{\textit{''Knowing your audience''} is easier said than done.} 
Reading order is shaped not only by design, but also by \thcontext{user expectations, domain familiarity, and task context}. Authors are aware of this complexity, and emphasize the need to understand their audience. However, the diversity of user flows is a testament to how challenging of a task this is. A comparison between author and user flows showed that author flows tended to be cleaner and more coherent, suggesting a tendency to underestimate the potential for diversity. As such, we posit that on top of knowing one's audience, it is also important to ``expect the unexpected'', by supporting flow diversity with a flexible and approachable design; whether that means understandable encodings, discoverable guidance, or allowing for multi-pronged exploration pathways.
% \implication{9}{Designing reading order with the \thcontext{audience and context} in mind is important, but expecting and designing for diversity is key. Providing navigation scaffolds and supporting multiple entry points and pathways can accommodate diverse reading strategies without enforcing a single rigid flow.}
\rr{\implication{9}{ Reading flow should be designed for structured variability rather than a single intended path. While understanding the \thcontext{target audience} remains important, designers should anticipate multiple coherent reading strategies by providing semantic anchors, navigation scaffolds, and flexible entry points that support diverse but interpretable pathways through the dashboard.}}

\subsection{Discussion}

% Reiterate patterns, and contrast with literature.
% - Literature that aligns and confirms our findings. Revisit dashboard design  recommendations for practitioners, for text design, from heuristics paper.

% - Comment on divergences against visual saliency models. \todo{assess if we have space to add those figures.} 
Taken together, our findings \rr{on a dashboard ``first contact'' task} suggest that dashboard reading order is best understood as a \emph{guided but underdetermined} process. Across dashboards, participants consistently relied on \thlayout{layout} as a first-pass scaffold: titles, KPIs, and overview elements served as common entry points, with flows often proceeding in a roughly top-down, left-to-right fashion when spatial organization and hierarchy were well aligned. This pattern reinforces established dashboard design guidance emphasizing clear information hierarchy, grouping, and overview-first organization~\cite{bach2022dashboard,sarikaya:2019}. It also aligns with empirical \rr{eye-tracking} evidence that layout order significantly shapes user attention and perceived complexity in dashboards~\cite{zhang2024effects}. Our findings extend these recommendations by showing that these structural cues are not merely aesthetic or organizational; they actively support users in constructing an initial mental model of the dashboard before engaging in more localized exploration.

The consistency of title- and overview-first behavior also aligns with prior work on the \thsemantic{semantic} and \thfunction{functional} roles of text in visualization. In our study, concise textual elements, especially titles, KPI headers, and short summaries were disproportionately important in establishing orientation, even in visually dense dashboards. This supports findings that text \rr{commands early attention~\cite{yang2025dashboard}}, plays a critical role in guiding interpretation and interaction~\cite{sultanum2024instruction}, and echoes recent work on dashboard text authoring that emphasizes its role in framing insights and connecting views~\cite{plume}. More broadly, this aligns with research in visualization accessibility and narrative design, which highlights how textual descriptions and summaries help structure user understanding~\cite{zong2022rich, Lundgard2021AccessibleVV}. However, our results refine these perspectives by showing that not all text contributes equally to reading flow: concise, high-signal text was frequently integrated into the primary reading path, whereas longer text blocks were often deferred or skipped unless tightly coupled to a visualization. This suggests that practitioners should prioritize brevity and tight visual-textual coupling, consistent with emerging heuristics for dashboard design~\cite{setlur2023heuristics}.

Our findings confirm prior work on multiview visualization by showing that users do not interpret dashboards as flat collections of independent components, but as \emph{\thlayout{\textit{locally}} \thsemantic{coherent} groups}. Participants frequently co-read titles with their associated charts, KPIs with nearby trends, and annotations with the views they described. This behavior reflects established models of understanding as an iterative process of information foraging and integration~\cite{pirolli2005sensemaking}, as well as prior work on coordinated multiple views that emphasizes the importance of spatial and semantic relationships between views~\cite{roberts:2007}. It also aligns with recent work showing that both structural and contextual relationships influence how users interpret collections of visualizations~\cite{chart2vec}. In this framing, reading order emerges from an interaction between global layout and local semantic neighborhoods: layout determines \emph{where to begin}, while proximity and meaning guide \emph{where to go next}. This helps explain why auxiliary elements (e.g., logos, metadata, controls) were consistently deprioritized; they are weakly connected to the analytic narrative despite being visually present.

At the same time, our results diverge from classical \thsaliency{visual saliency} models in important ways. While visually salient elements, such as saturated colors, large marks, or distinctive chart types, occasionally attracted early attention, they did not reliably determine reading order. Instead, semantically meaningful elements (e.g., titles, KPIs) were often prioritized even when they were less visually prominent. This contrasts with traditional saliency-based accounts of visual attention~\cite{itti:1998}, and supports prior findings that such models are insufficient for predicting attention in data visualizations because they fail to account for \thsemantic{semantic features} like text \cre{content}~\cite{matzen:2017,yang2025dashboard}. Similarly, common scanning patterns such as the F-shaped reading pattern~\cite{nielsen2006fpattern} only partially explain dashboard traversal: while some initial movements followed these patterns, subsequent transitions were heavily shaped by \thcontext{personal relevance} and interpretive styles.\,Our findings therefore suggest that reading order in dashboards is not purely stimulus-driven, but reflects an interaction between perceptual saliency, semantic meaning, and user intent.

More broadly, the variability we observed, particularly in mid-flow transitions and around \thinteract{interactive} or \thfunction{peripheral elements}, suggests that dashboards do not impose a single, fixed reading path. This aligns with conceptualizations of dashboards as flexible information spaces supporting multiple analytical trajectories \cite{benyon:2001, sarikaya:2019}. However, this variability was structured rather than arbitrary: participants showed strong agreement on entry points and local groupings, diverging mainly when multiple plausible next steps existed. This suggests an important implication: rather than enforcing a rigid sequence, practitioners should aim to make a \emph{preferred path} legible by aligning \thlayout{layout}, \thsemantic{concise text}, and semantic grouping, while using \thsaliency{visual saliency} or unconventional placement deliberately to challenge or redirect that path.

\rr{Finally, although we focused on reading order for dashboards, our learnings could potentially extend to other multi-view visualization artifacts such as analytical reports~\cite{ferilli2014abstract,wang2021layoutreader}, data stories~\cite{segel2010narrative}, and visualization atlases~\cite{wang2025visualization}. 
%Our findings suggest that dashboard design factors such as hierarchy, semantic importance, visual prominence, and interaction affordances influence not only which elements attract attention, but also how users sequence their interpretation of them. 
Understanding these reading flows may support the design of more navigable and interpretable visualization systems, including tools to explicitly guide attention, communicate intended analytical pathways, or adapt to user-specific exploration strategies.}

%% file: tex/06-reflections.tex
\section{Research Opportunities and Applications}
\label{sec:reflections}
Our findings reframe dashboard reading order as a first-class design consideration, rather than an implicit byproduct of layout and visual attention alone. While prior work has focused on layout, encoding, and interaction as the primary levers of dashboard design~\cite{sarikaya:2019,bach2022dashboard}, \rr{our results suggest that reading flow provides an additional lens for understanding and designing dashboard experiences.} By making reading flow explicit, we open new opportunities for design practice and system support that better align with how users traverse dashboards.

\pheading{Reading order authoring support.}
Our findings suggest that designers rarely encode reading order explicitly; instead, flow emerges implicitly from layout and content decisions. \rr{Current authoring tools primarily support visual composition rather than intended reading flow or semantic structure~\cite{sultanum2024instruction,plume}. Future systems could predict likely reading flows, visualize alternative traversal paths, or recommend layout refinements based on communication goals and target audiences. This direction complements prior work on design recommendation and layout adaptation~\cite{lin:2024,zeng2023semi}, while extending document reading-order models~\cite{ferilli2014abstract,wang2021layoutreader} to interactive dashboards.} %This direction builds on prior work in design mining and recommendation systems~\cite{lin:2024}, as well as semi-automatic layout adaptation~\cite{zeng2023semi}. It also connects to research on reading order detection in documents~\cite{ferilli2014abstract, wang2021layoutreader}, though dashboards introduce additional complexity due to interaction, heterogeneous views, and non-linear navigation. Supporting designers in exploring multiple plausible pathways remains a key opportunity.

\pheading{Reader support and personalization.}
While aggregate patterns reveal shared tendencies (e.g., prioritizing titles and overview elements), we also observe substantial variability across users. Individuals often exhibit consistent strategies across dashboards, yet differ significantly from one another when interacting with the same design. This aligns with prior work showing that user characteristics and expertise influence attention and interpretation~\cite{toker:2013, wallace2022towards}. This variability suggests \rr{opportunities for personalized navigation support, adapting reading guidance to user expertise, goals, or prior behavior~\cite{dashguide}}. Another promising direction is dynamically modulating visual emphasis: highlighting overlooked but important elements or deemphasizing already-visited components. Prior work on attention guidance~\cite{joshi2024constrained} suggests that such cues can steer behavior without enforcing rigid sequences.

\pheading{Accessibility and alternative representations.}
Reading order also has implications for accessibility. For users relying on screen readers or assistive technologies, dashboards must be translated into linear or structured representations that preserve meaning and context. \rr{Existing approaches translate dashboards into linear representations for screen readers~\cite{zong2022rich,srinivasan2023azimuth,Lundgard2021AccessibleVV}, but often lack information about meaningful traversal sequences. Explicit reading-flow models could preserve semantic groupings and transitions when serializing dashboards, enabling richer textual summaries, guided narratives, and hierarchical navigation~\cite{soto2019visual,zhang2020trie}}. More broadly, this enables alternative representations such as textual summaries, guided narratives, or hierarchical navigation structures, supporting a wider range of users and interaction modalities.

\pheading{Semantic layout adaptations and responsive design.} \rr{Our findings also motivate reading-flow-aware responsive design. Existing responsive techniques primarily resize and rearrange components~\cite{zeng2023semi}. In contrast, reading flow provides a behavioral objective for adaptation: preserving meaningful entry points, groupings, and transitions across devices rather than simply maintaining spatial layouts. This perspective complements emerging work on structure-aware visualization and layout adaptation~\cite{listructureaware:2022,chart2vec}}.
%Our findings also point to opportunities for adapting dashboard layouts based on reading order, particularly in constrained environments such as mobile devices. Existing responsive design approaches primarily focus on resizing and rearranging components~\cite{zeng2023semi}, often prioritizing visual constraints over semantic relationships. In contrast, reading order provides a principled basis for adaptation. Rather than simply reflowing elements, systems could reorganize dashboards to preserve meaningful sequences and groupings, ensuring that key entry points and transitions remain intact. This aligns with emerging work on structure-aware visualization and layout adaptation~\cite{listructureaware:2022, chart2vec}, and reframes responsive design as maintaining coherent information pathways across contexts.

Taken together, these opportunities shift the view of dashboards from static visual compositions to dynamic information spaces shaped by user traversal~\cite{benyon:2001,pirolli2005sensemaking}. By explicitly modeling and supporting reading order, future systems can better align design intent with user experience, enabling more effective, adaptive, and inclusive dashboard interactions.

%% file: tex/07-limitations.tex
\section{Limitations}
Our findings should be interpreted in light of several limitations related to both the study design and the participant sample.
First, our participant pool was skewed toward business-oriented users who regularly engage with KPI-driven and monitoring-style dashboards. This likely introduced a bias toward familiar patterns such as top-down, left-to-right flows, and a \rr{focus on} summary metrics.
%, and emphasis on performance tracking. 
While we included a diverse set of dashboards, including infographic-style and public-facing examples, prior participant experience likely influenced how they interpreted and traversed these examples. Future work with broader populations
%, including novice users and general audiences, 
may reveal alternative reading strategies, particularly for 
%less structured or 
more narrative-driven dashboards.

Second, \rr{our investigation is limited by task, flow, and dashboard variety, which impacts the generalizability of our findings}. The study focused on initial understanding and onboarding, which may not reflect real-world usage. In practice, users approach dashboards with specific goals, time constraints, and prior knowledge, all of which shape traversal patterns \rr{in non-linear ways}. Our findings therefore capture an early-stage mode of engagement \rr{and reading order characterization} rather than the full spectrum of use. \rr{While diverse, our dashboard stimuli set is small and constrained by our author volunteer pool. Future studies may consider more fine-grained control of dashboard features, more diverse flow features, and a broader set of user tasks and intents.}

Third, while we analyzed aggregate patterns across participants, reading order remains inherently variable and context-dependent. As the notion of a single ``ground truth'' reading order \rr{appears unlikely}, our results characterize distributions of plausible flows. Although we report areas of convergence and divergence, the space of valid reading flows is broader than what can be captured in controlled study settings.

Finally, our study relies on self-reported and externally articulated reading sequences rather than direct behavioral measures such as eye tracking or interaction logs. While this approach provides rich insight into participant reasoning and intent, it may not fully capture subconscious attention patterns or fine-grained navigation behaviors.

%% file: tex/08-conclusion.tex
\section{Conclusion}
In this work, we conceptualize dashboard reading order not as a fixed sequence, but as a dynamic process shaped by the interplay of layout, visual saliency, semantics, functional roles, interaction, and user context. \rr{While users often share common entry points and structural tendencies, their reading flows diverge as they pursue different goals, respond to different cues, and draw on varying levels of familiarity. As Sir John Lubbock, a 19th-century naturalist observed, ``\textit{What we see depends mainly on what we look for}''~\cite{lubbock1894}. In the context of dashboards, what users seek, whether a specific metric, a trend, or a broader narrative, fundamentally shapes how they traverse and interpret information. Viewing reading flow as an adaptive process rather than a fixed property opens opportunities for authoring tools, accessibility support, and agentic analytics that anticipate multiple valid reading flows and better align dashboard design experiences with diverse user needs.} %Our findings show that while users often share common entry points and structural tendencies, their paths diverge as they prioritize different elements, respond to different cues, and adapt to their goals and familiarity. Reading flows, therefore, are not prescribed by design alone; they are co-constructed through perception, intent, and use. This perspective is aptly captured by Sir John Lubbock, a 19th-century naturalist: ``\textit{What we see depends mainly on what we look for}''~\cite{lubbock1894}. In the context of dashboards, what users look for, whether a specific metric, a trend, or a broader narrative, fundamentally shapes how they traverse and interpret the interface. Recognizing this variability opens the door to more adaptive and user-aligned systems: from authoring tools that anticipate multiple valid flows, to assistive technologies that guide users through context-aware pathways, to agentic analytics that proactively surface relevant information. Together, these directions point toward a future where dashboards are not only designed for presentation, but actively support diverse, evolving modes of data understanding.

%% file: tex/09-appendix.tex
To complement the aggregate findings presented in the paper and further contextualize fine-grained study findings and implications, we provide the full list of individual reading flow instances provided by both authors and end-users. Flow instances are rendered using a multi-hue \texttt{rainbow} colormap\footnote{\url{https://matplotlib.org/stable/users/explain/colors/colormaps.html}} to clearly emphasize flow stage transitions, spanning 
{\color[HTML]{F0311E}\textbf{early flow}}
\includegraphics[width=5em, height=.77em]{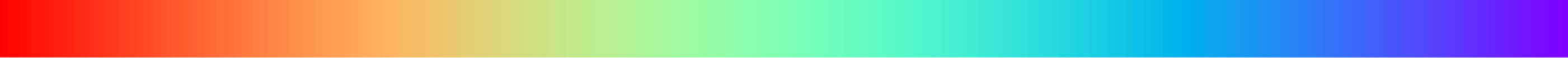}
{\color[HTML]{692BF4}\textbf{later flow}} through {\color[HTML]{73e1ab}\textbf{mid-flow}} components. %7FF7BB
%\textit{early visited components in {\color{red}{red}} 
%to later visited ones in {\color{blue}{blue}}}. 
Skipped components are not factored into the flow, and are rendered with a red {\color{red}{X}} (e.g., the ``Website'' button in both author flow instances for D1 \autoref{fig:appendix-author-flows-1}). They are also accompanied by short summaries of flow variations, underscoring the diversity across flow instances and dashboards.

\subsection{Author Reading Flows}
\label{sec:author_flows}

\begin{figure*}[t]
    \centering

    \begin{subfigure}{\linewidth}
        \centering
        \includegraphics[width=0.8\linewidth]{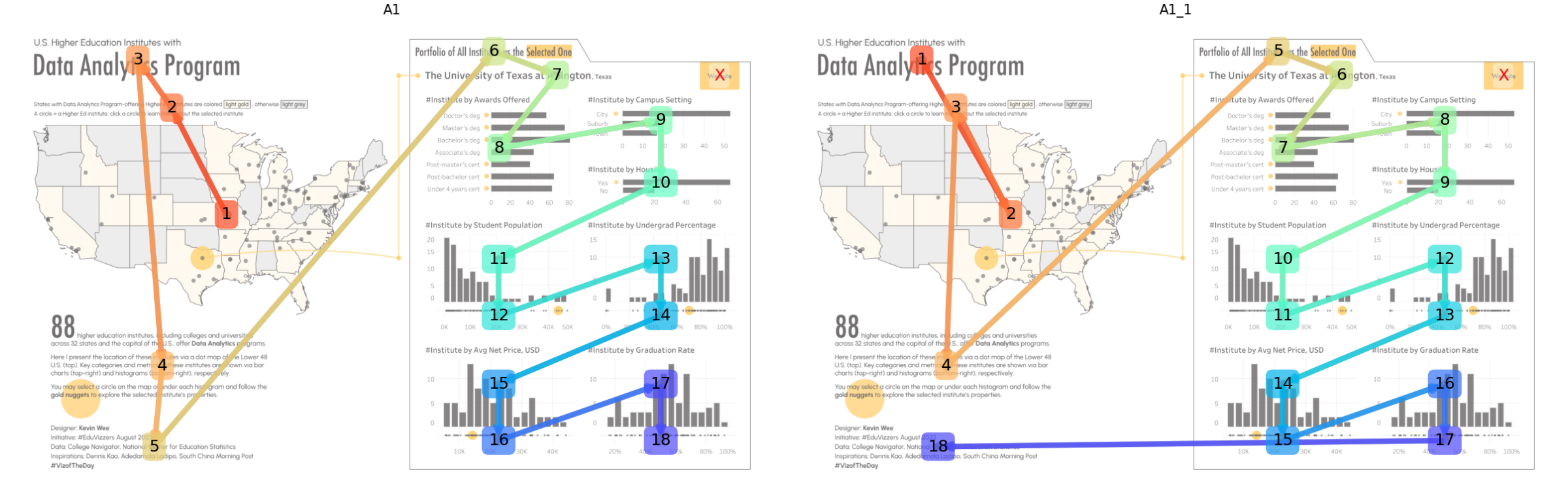}
        \caption{\textbf{A1:} \textbf{(left)} Logical flow anchored by the map as the primary visual entry point, designed to attract attention and introduce geographic context, with supporting title and legend framing interpretation. \textbf{(right)} A `mobile screen' serialized flow, where reading order is structured linearly, beginning with the title, followed by the map and supporting text, and progressing top-to-bottom through the dashboard.} 
    \end{subfigure}

    \vspace{0.5em}

    \begin{subfigure}{\linewidth}
        \centering
        \includegraphics[width=0.8\linewidth]{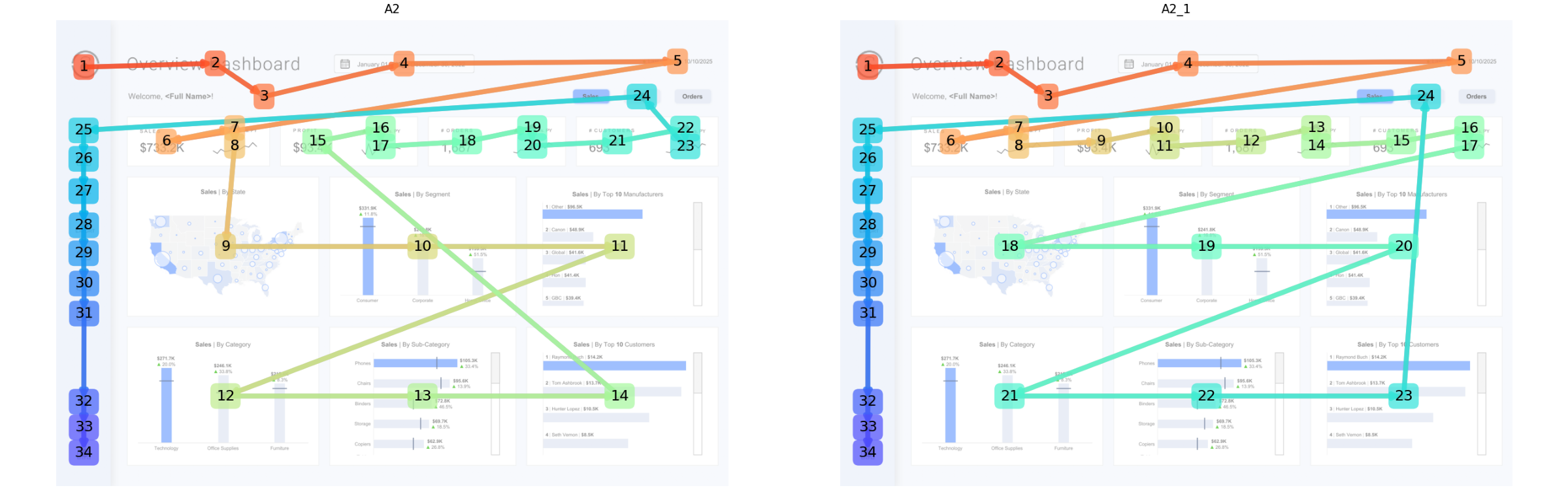}
        \caption{\textbf{A2:} \textbf{(left)} Logical flow guided by layout and semantics, beginning with top-left contextual elements (title, dates, and metadata), followed by high-level KPIs and then detailed charts in a zig-zag progression. \textbf{(right)} KPI-first flow reflecting the original design intent, where users prioritize top-level metrics (BANs) before transitioning to detailed charts, followed by interaction elements such as filters and export.}
    \end{subfigure}

    \vspace{0.5em}

    \begin{subfigure}{\linewidth}
        \centering
        \includegraphics[width=0.3\linewidth]{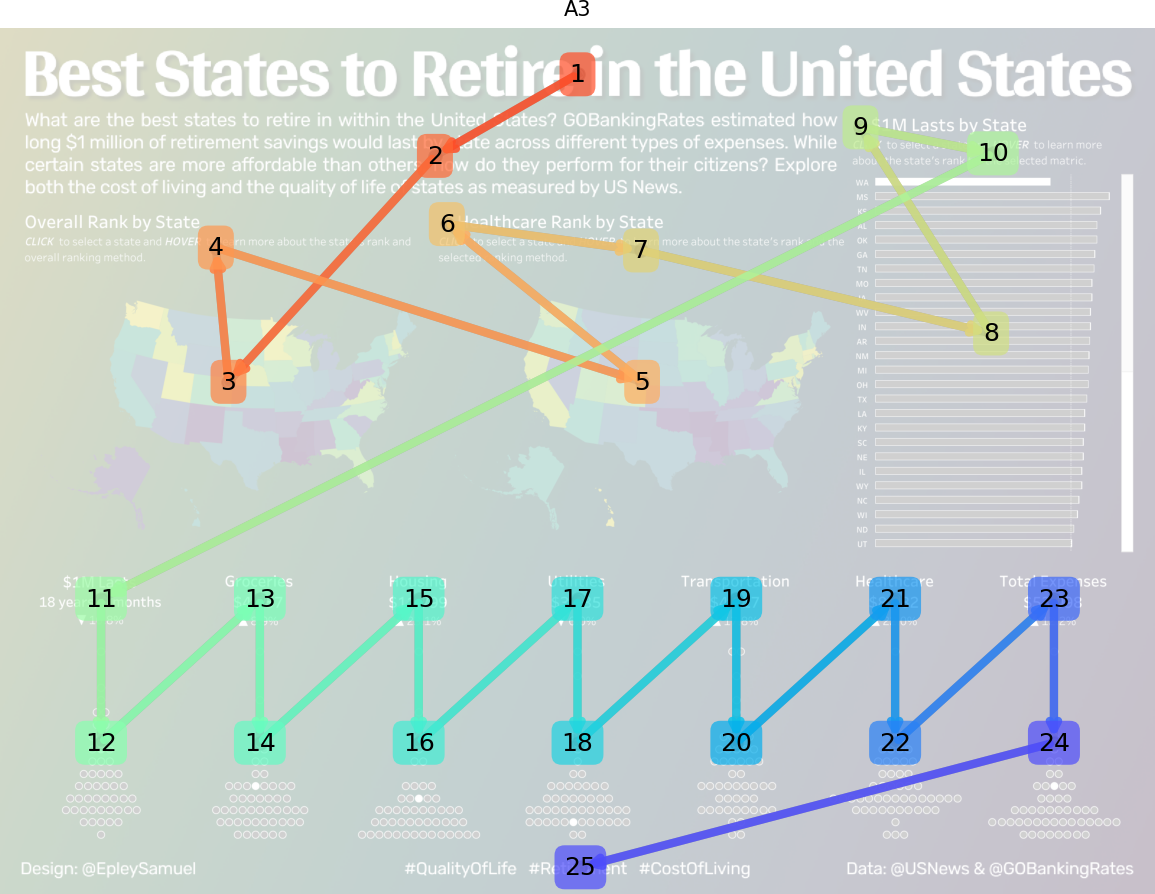}
        \caption{\textbf{A3:} Open, exploration-driven flow that begins with title and overview maps, followed by interactive comparisons and summary metrics, supporting flexible, non-linear traversal rather than a fixed sequence.}
    \end{subfigure}

    \vspace{0.5em}

    \begin{subfigure}{\linewidth}
        \centering
        \includegraphics[width=0.9\linewidth]{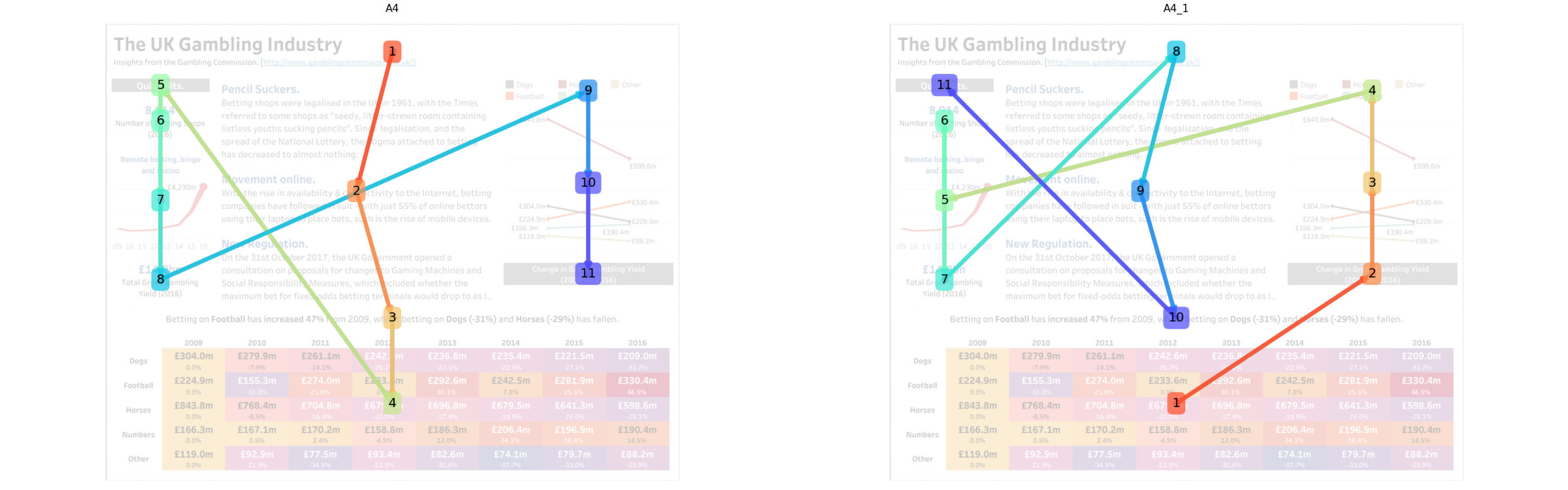}
        \caption{\textbf{A4:} \textbf{(left)} Narrative-driven ``newspaper'' flow, where reading begins with title and text to establish the story, followed by supporting charts and figures, with legends and auxiliary elements deferred. \textbf{(right)} Visual saliency-driven flow, where attention is captured by high-contrast and visually dense elements (e.g., heatmap and slope chart), with text skimmed and interpretation guided by perceptual prominence rather than narrative order.}
    \end{subfigure}

    \caption{
    Author-specified reading flows \textbf{(A1–A4)}, illustrating diverse traversal strategies including saliency-driven entry (A1), layout- and semantics-guided progression (A2), open exploration (A3), and narrative versus saliency-driven divergence (A4).}
    \label{fig:appendix-author-flows-1}
\end{figure*}

\begin{figure*}[t]
    \centering

    \begin{subfigure}{\linewidth}
        \centering
        \includegraphics[width=0.9\linewidth]{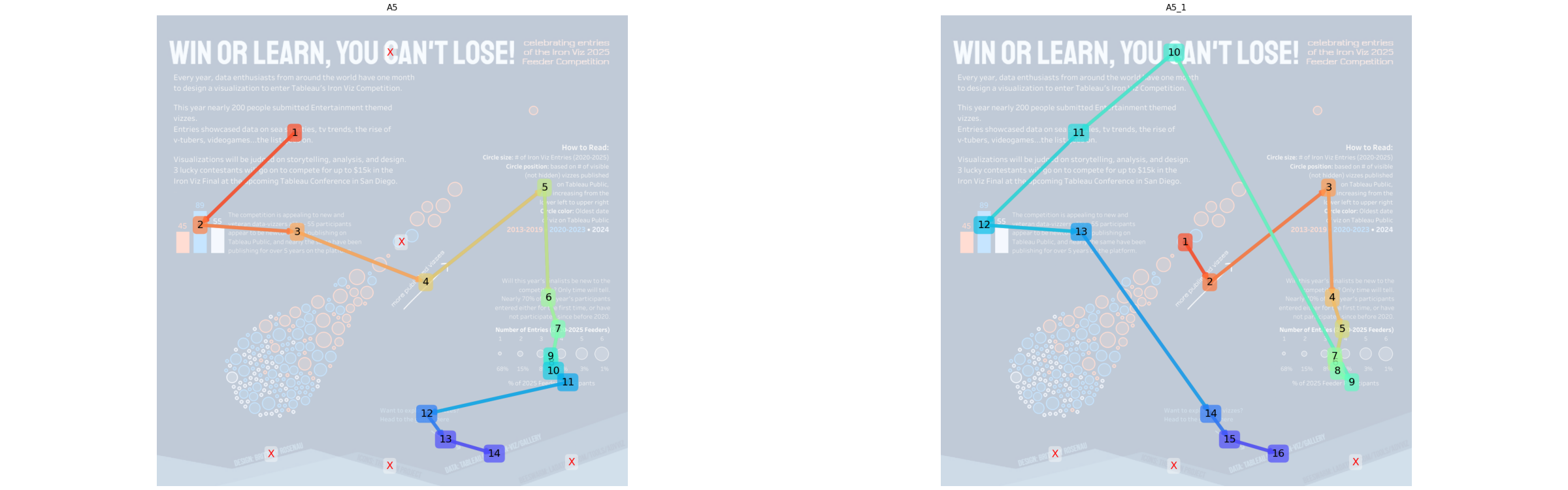}
        \caption{\textbf{A5:} \textbf{(left)} Infographic-style flow anchored by a visually striking beeswarm and title as dual entry points, followed by contextual text and guided exploration, with annotations and legends supporting progressive understanding and encouraging users to return to the main visualization. \textbf{(right)} Visual-first flow where attention begins with the beeswarm, followed by transitions to legends and context, reflecting non-linear traversal that prioritizes saliency over narrative sequence.}
    \end{subfigure}

    \vspace{0.5em}

    \begin{subfigure}{\linewidth}
        \centering
        \includegraphics[width=0.8\linewidth]{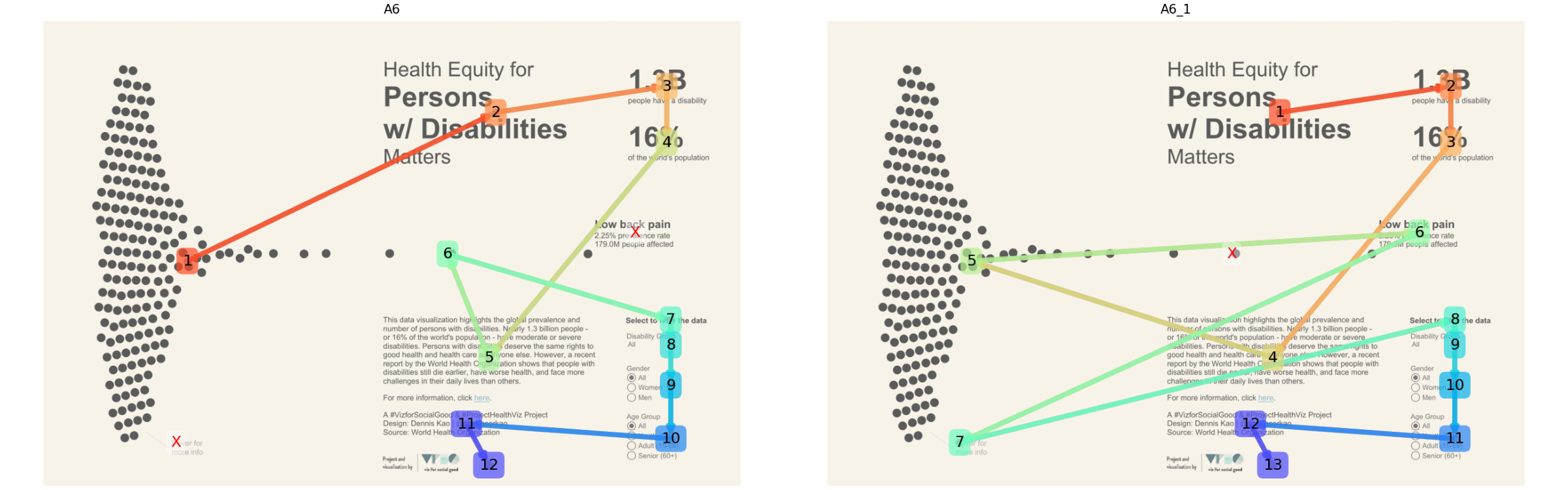}
        \caption{\textbf{A6:} \textbf{(left)} Visual-first exploratory flow in which attention is drawn to the beeswarm as a salient entry point, followed by movement to title, summary metrics, and descriptive text to establish context before deeper inspection of the data. \textbf{(right)} Accessibility-driven serialized flow, where reading order is linearized to begin with title and summary metrics, followed by descriptive text and structured data, prioritizing clarity over visual exploration.}
    \end{subfigure}

    \vspace{0.5em}

    \begin{subfigure}{\linewidth}
        \centering
        \includegraphics[width=0.75\linewidth]{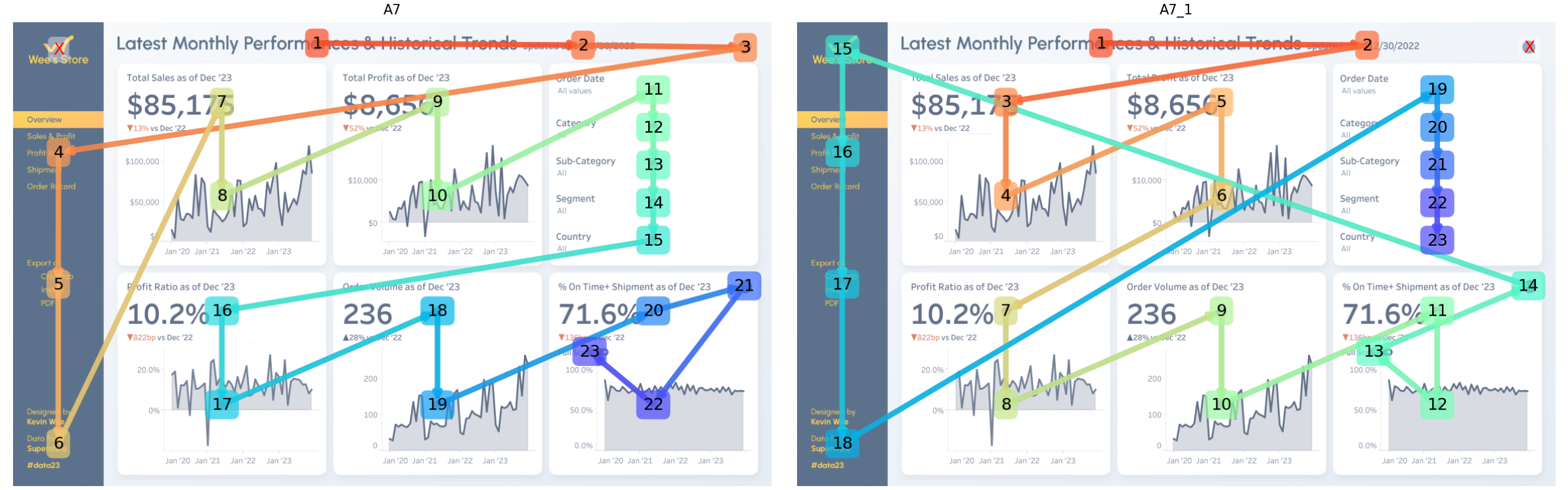}
        \caption{\textbf{A7:} \textbf{(left)} First-time onboarding flow in which users begin with title and guidance elements (e.g., info panel and menus), followed by high-level summary metrics and charts, with interaction and filters explored after initial orientation. \textbf{(right)} Mobile-optimized flow prioritizing immediate access to key metrics, with top-down progression through KPI–chart pairs and interaction elements (navigation and filters) consolidated into persistent bottom controls.}
    \end{subfigure}

    \vspace{0.5em}

    \begin{subfigure}{\linewidth}
        \centering
        \includegraphics[width=0.8\linewidth]{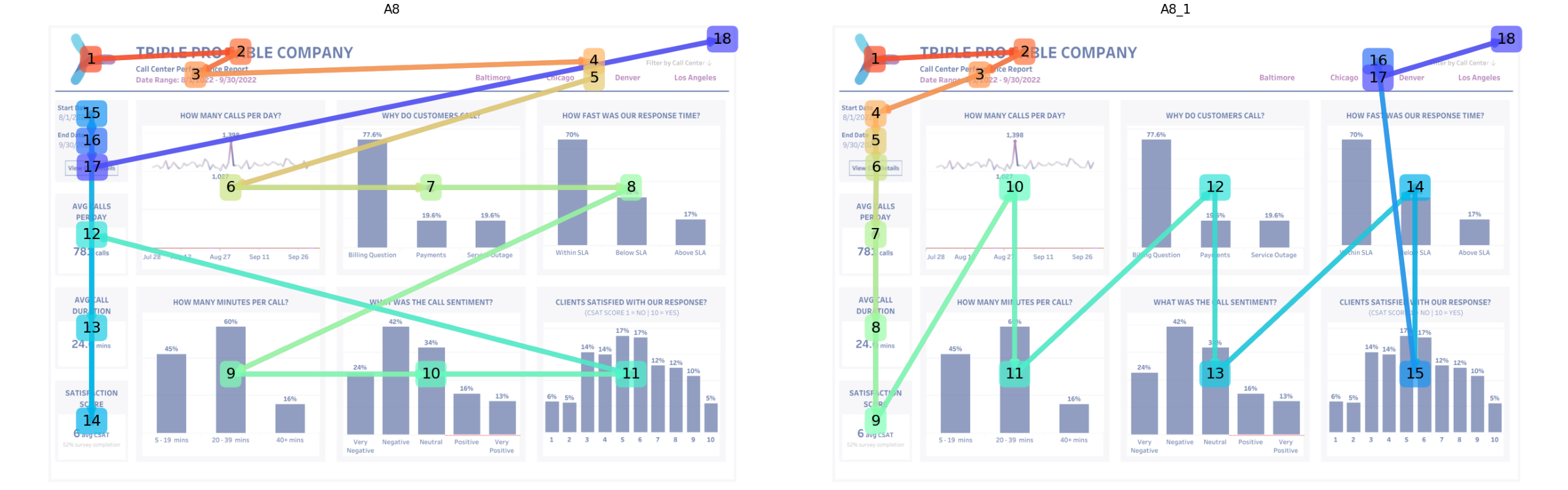}
        \caption{\textbf{A8:} \textbf{(left)} Header-first flow reflecting design intent, where users scan the top row for orientation, then prioritize primary charts for overview and progressively move to more detailed views, with sidebar KPIs and interaction elements treated as secondary. \textbf{(right)} KPI-first flow in which users begin with left-side summary metrics to establish a high-level understanding before transitioning to detailed charts, reflecting a pragmatic, summary-driven reading strategy.}
    \end{subfigure}

   \caption{
Author-specified reading flows \textbf{(A5–A8)}, illustrating diverse traversal strategies including curiosity-driven infographic flow (A5), visual-first and accessibility-driven divergence (A6), onboarding vs. mobile interaction patterns (A7), and contrasts between layout-driven and KPI-first reading (A8).}
    \label{fig:appendix-author-flows-2}
\end{figure*}

Author-generated reading flows (\pid{A1}–\pid{A8}) reflect intended traversal strategies, capturing how designers expected (or intended) users to navigate dashboard content given the design choices they made. These flows are shaped by design goals, layout decisions, and assumptions about user behavior, and often emphasize structured progressions such as overview-to-detail, narrative-driven sequences, or exploratory engagement. Across dashboards, authors frequently encode semantic hierarchy through layout, prioritizing key entry points such as titles, summaries, or visually salient elements before transitioning to more detailed views. At the same time, authors anticipated multiple possible traversal paths, which mapped to alternative pathways that account for different user entry points, levels of familiarity, device constraints, or accessibility considerations. As such, most authors chose to define \textit{alternative} flow instances to capture such variants.
Figures~\ref{fig:appendix-author-flows-1} and~\ref{fig:appendix-author-flows-2} illustrate these author-envisioned flows across all dashboards.

\subsection{End-User Reading Flows}
\label{sec:enduser_flows}

\begin{figure*}[t]
    \centering

    \begin{subfigure}{\linewidth}
        \centering
        \includegraphics[width=0.82\linewidth]{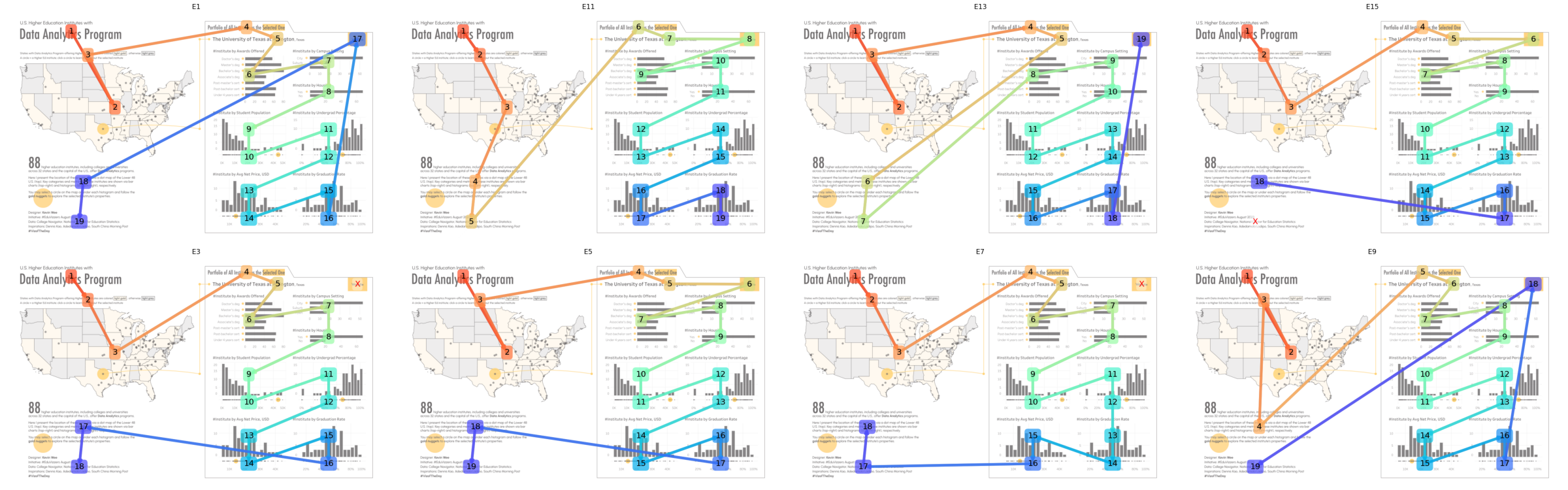}
        \caption{\textbf{D1:} End-users consistently begin with the title and map, then transition to the right-hand panel via the visual connector, generally scanning charts left-to-right and top-to-bottom; however, frequent back-and-forth between the map and charts reflects interaction-driven exploration, while supporting text, metadata, and the website link are typically deferred.}
    \end{subfigure}

    \vspace{0.75em}

    \begin{subfigure}{\linewidth}
        \centering
        \includegraphics[width=0.83\linewidth]{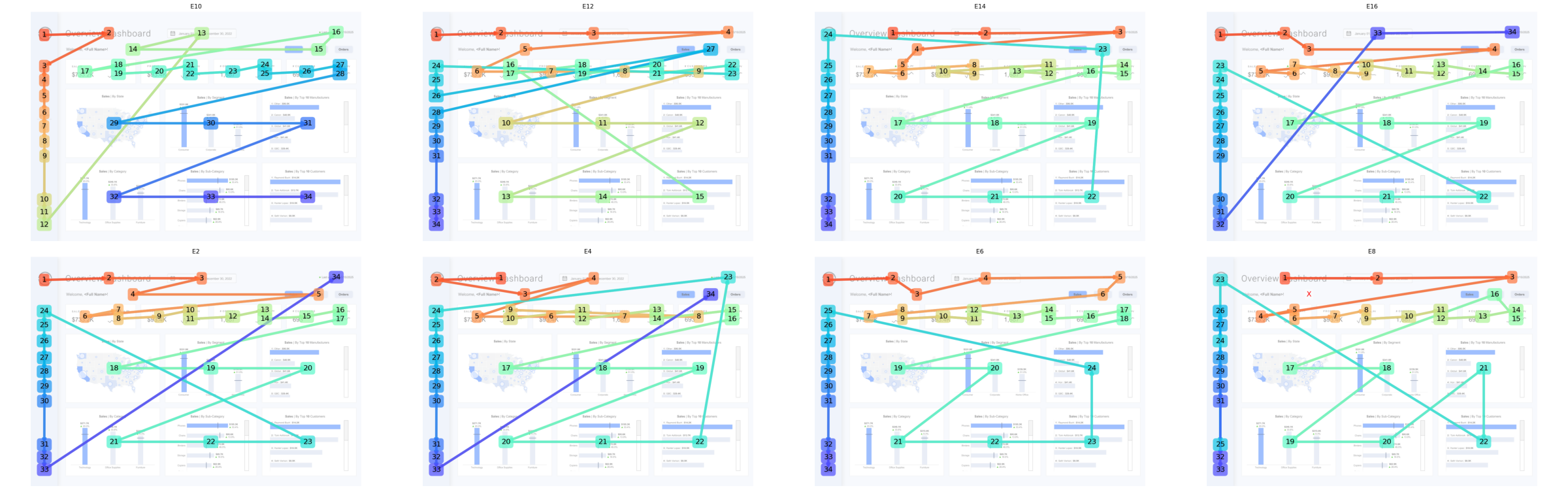}
        \caption{\textbf{D2:} End-users follow a structured top-down flow, starting with title, filters, and KPIs, then scanning charts left-to-right while frequently looping through tabs and filters to compare views; interaction and re-evaluation are central to navigation, with metadata and controls addressed last.}
    \end{subfigure}

    \vspace{0.75em}

    \begin{subfigure}{\linewidth}
        \centering
        \includegraphics[width=0.85\linewidth]{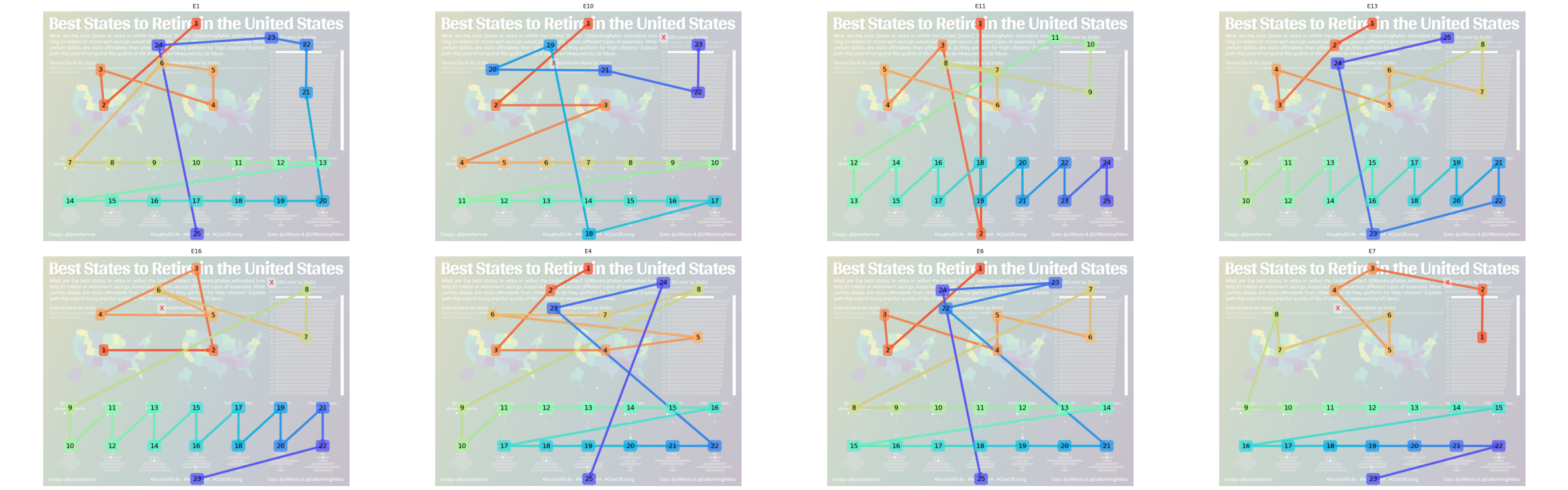}
        \caption{\textbf{D3:} End-users generally follow a top-down flow starting with title and maps, then moving to KPIs and charts, while frequently revisiting maps and text to build understanding; unfamiliar visuals (e.g., beeswarms) are deferred or reinterpreted after contextual cues, with filters and metadata explored last.}
    \end{subfigure}

    \vspace{0.75em}

    \begin{subfigure}{\linewidth}
        \centering
        \includegraphics[width=0.88\linewidth]{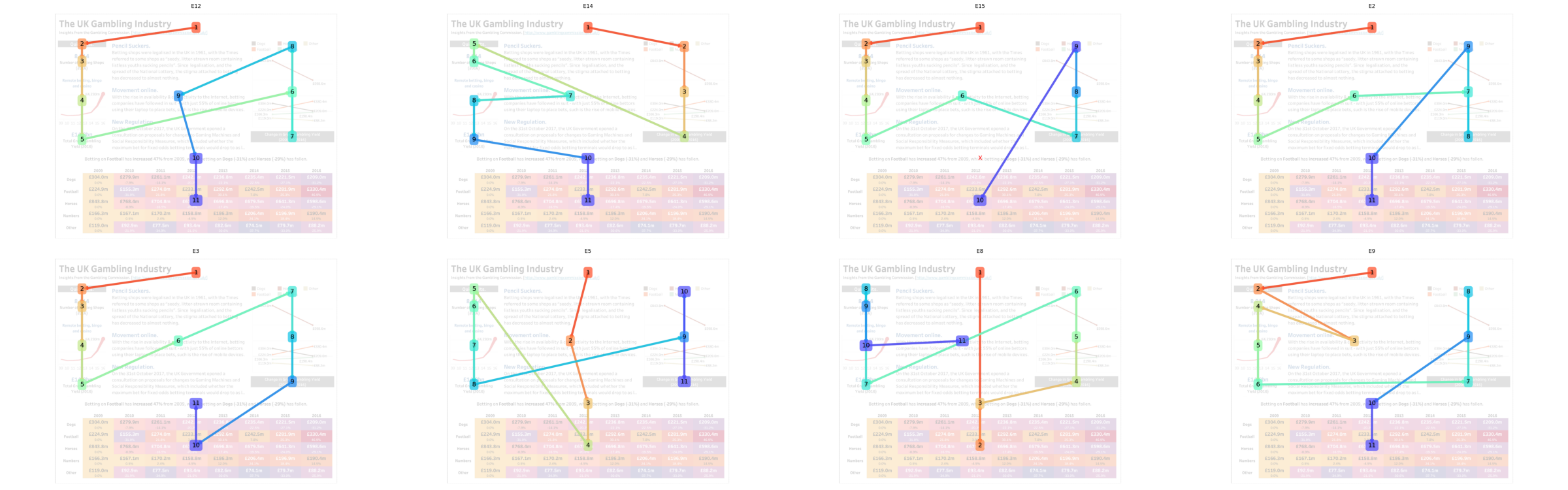}
        \caption{\textbf{D4:} User flows are heterogeneous; some follow a text-first, top-down narrative, while others jump to visually salient charts; navigation often alternates between text and visuals to resolve ambiguity, with dense text and less intuitive elements (e.g., heatmap) deferred or revisited later.}
    \end{subfigure}

\caption{
End-user reading flows for dashboards (\db{1}–\db{4}). Each subfigure overlays multiple users’ scanpaths, revealing consistent entry points (e.g., titles and primary visuals) alongside variability in navigation strategies, including top-down scanning, visual-first exploration, and frequent back-and-forth between views to build understanding.
}
    \label{fig:appendix-db-user-flows-1}
\end{figure*}

\begin{figure*}[t]
    \centering

    \begin{subfigure}{\linewidth}
        \centering
        \includegraphics[width=0.93\linewidth]{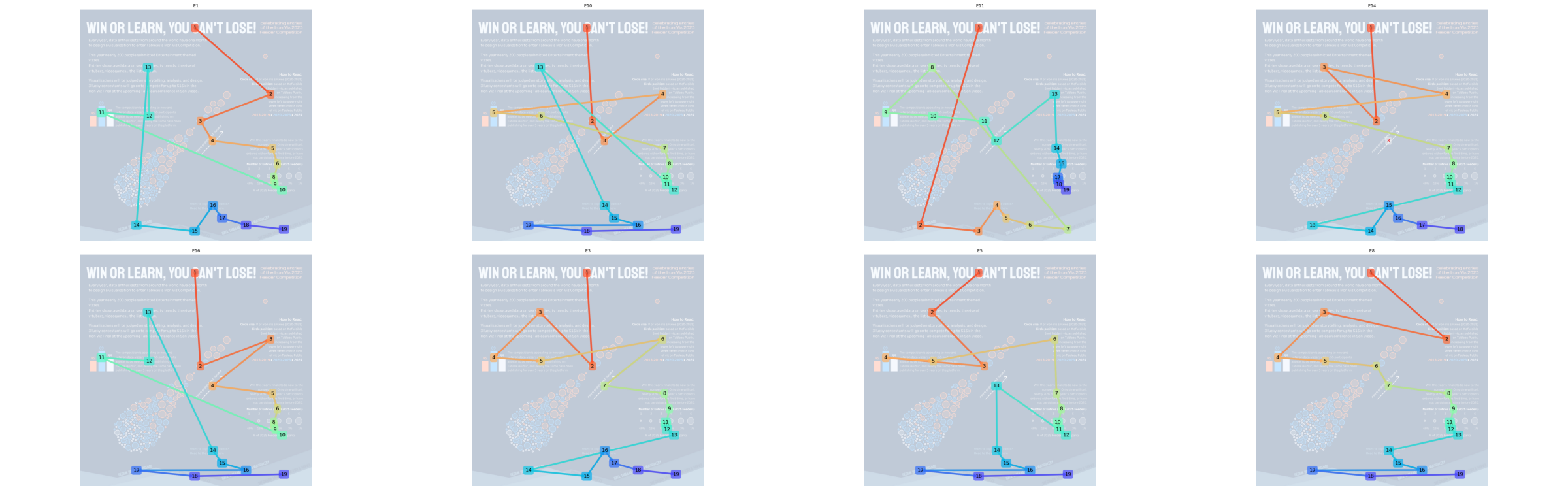}
        \caption{\textbf{D5:} Users show highly non-linear, visually driven behavior, often jumping between the beeswarm and explanatory elements (e.g., ``How to read''), with frequent back-and-forth to resolve interpretation; text is commonly deferred until needed, and understanding emerges through iterative exploration rather than a fixed sequence.}
    \end{subfigure}

    \vspace{0.75em}

    \begin{subfigure}{\linewidth}
        \centering
        \includegraphics[width=0.85\linewidth]{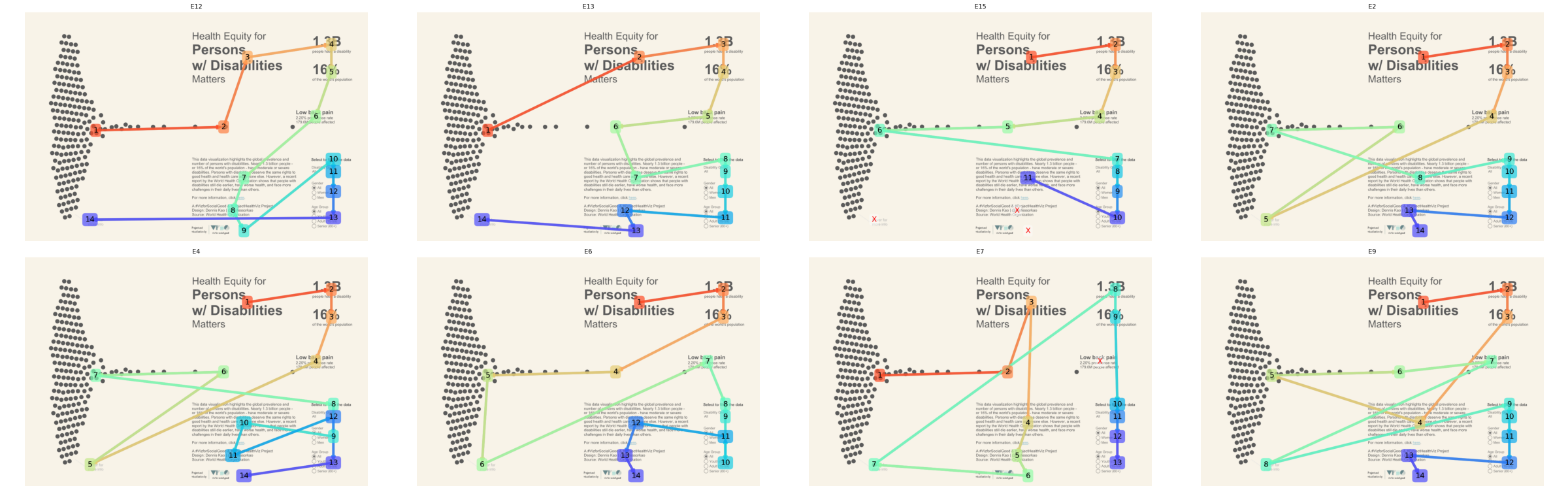}
        \caption{\textbf{D6:} Users exhibit mixed entry points; some start with title and KPIs, others with the beeswarm, often focusing on outliers first; navigation is highly interactive, with frequent filtering and revisiting of charts to compare changes, while explanatory text and metadata are typically secondary.}
    \end{subfigure}

    \vspace{0.75em}

    \begin{subfigure}{\linewidth}
        \centering
        \includegraphics[width=0.83\linewidth]{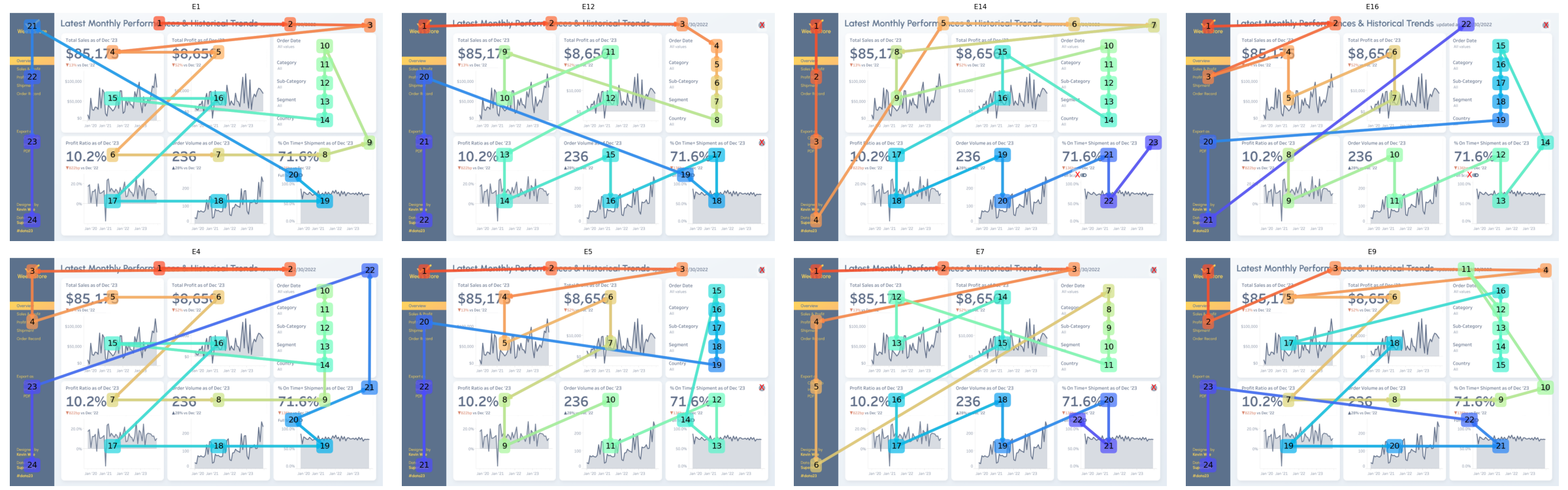}
        \caption{\textbf{D7:} Users largely follow a structured overview-first pattern starting with title and KPIs to build context, then exploring charts and filters iteratively; navigation is highly interactive, with frequent back-and-forth between metrics, filters, and tabs, while side panels and metadata are typically secondary.}
    \end{subfigure}

    \vspace{0.75em}

    \begin{subfigure}{\linewidth}
        \centering
        \includegraphics[width=0.86\linewidth]{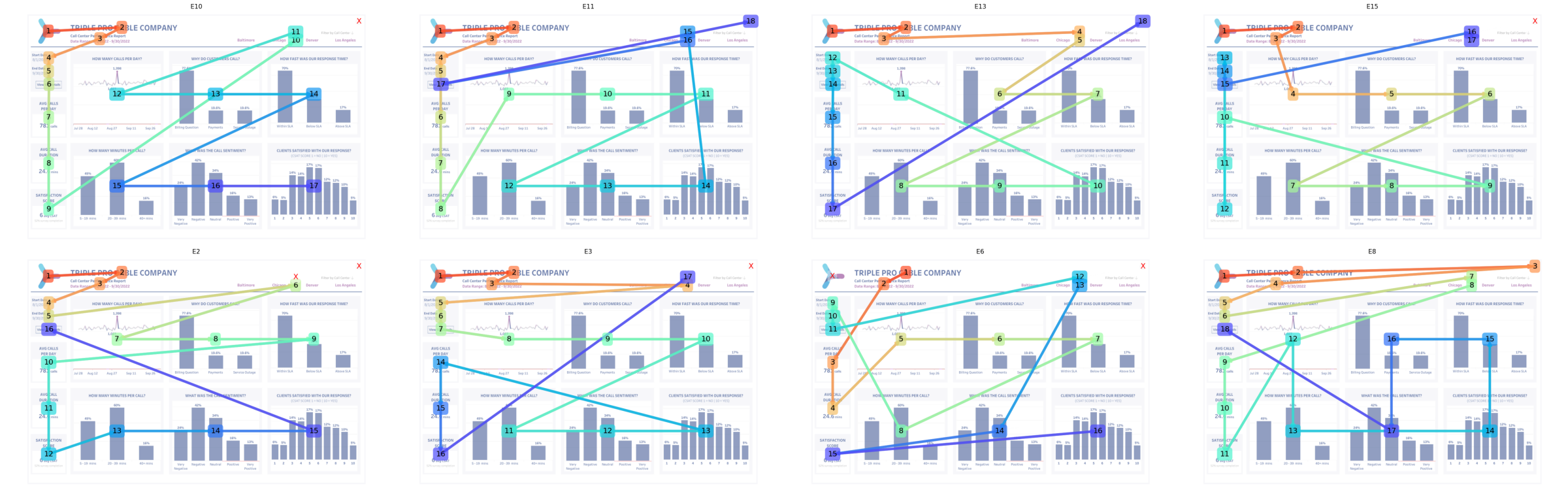}
        \caption{\textbf{D8:} Users begin with title and temporal context, then follow a largely left-to-right scan of charts, often prioritizing main visuals over side KPIs; filters and location tabs are explored iteratively, with frequent back-and-forth to contextualize metrics, while secondary elements (info, metadata) are typically ignored.}
    \end{subfigure}

\caption{
End-user reading flows for dashboards \textbf{(\db{5}–\db{8})}. Each subfigure overlays multiple users’ flows, highlighting visually driven entry points (e.g., prominent charts or titles), iterative navigation between context and detail (e.g., filters, legends, and KPIs), and back-and-forth exploration as users refine understanding, with secondary elements (e.g., metadata or auxiliary controls) often deferred or ignored.
}
    \label{fig:appendix-db-user-flows-2}
\end{figure*}

End-user reading flows captured how participants reasoned through ``logical'' reading flows, which reflect a combination of personal sensemaking strategies, visual traversal preferences (e.g., more layout-oriented, or saliency-leaning), and level of interest. They are stratified by dashboard, encompassing 8 flow instances each: \autoref{fig:appendix-db-user-flows-1} for \db{1}-\db{4}, and \autoref{fig:appendix-db-user-flows-2} for \db{5}-\db{8}. For each dashboard, we provide a short summary of collective flow behavior. Overall, they clearly showcase how no two flows are the same; numerous partial similarities emerge upon careful scrutiny.

